\documentclass{sig-alternate-10pt}
\usepackage{float}
\usepackage{time}
\usepackage{multirow}
\usepackage{multicol}
\usepackage{cite}
\usepackage[english]{babel}

\usepackage[font={small,bf}]{caption}
\usepackage{color}

\usepackage[caption=false]{subfig}
\usepackage{float}

\usepackage{graphicx}
\usepackage[hyphens]{url}

\newcommand{\todo}[1]{\textcolor{red}{TODO: #1}}

\begin{document}

\title{Reverse Engineering Socialbot Infiltration \\Strategies in Twitter}

\numberofauthors{3} 
\author{
\alignauthor Carlos A. Freitas\\
        \affaddr{UFMG, Belo Horizonte, Brazil}\\
        \email{alessandro@dcc.ufmg.br}	
\alignauthor Fabricio Benevenuto\\
        \affaddr{UFMG, Belo Horizonte, Brazil}\\
        \email{fabricio@dcc.ufmg.br}
 \and
 \alignauthor Saptarshi Ghosh\\
        \affaddr{IIEST Shibpur, India}\\
        \email{sghosh@cs.becs.ac.in}
 \alignauthor Adriano Veloso\\
        \affaddr{UFMG, Belo Horizonte, Brazil}\\
        \email{adrianov@dcc.ufmg.br}
}


\setcounter{table}{0}

\maketitle

\abstract{
\noindent 
Data extracted from social networks like Twitter are increasingly 
being used to build
applications and services that mine and summarize public reactions to events,
such as traffic monitoring platforms, identification of epidemic outbreaks, and 
public perception about people and brands.  
However, such services are vulnerable to attacks
from socialbots $-$ automated accounts that mimic real users $-$ 
seeking to tamper statistics
by posting messages generated automatically and
interacting with legitimate users. 
Potentially, if created in large scale, socialbots could be used to bias or 
even invalidate many existing services, by infiltrating the social
networks and acquiring trust of other users with time.
This study aims at understanding infiltration strategies of socialbots in the
Twitter microblogging platform.  To this end, we create 120 socialbot accounts with
different characteristics and strategies (e.g., gender specified in the profile, how active they are, the method used to generate their tweets, and the group of users they interact with), and
investigate the extent to which these bots are able to infiltrate the Twitter social network.  
Our results show that even socialbots employing simple automated mechanisms are able
to successfully infiltrate the network. Additionally, using a $2^k$ factorial design, we quantify infiltration effectiveness of different bot strategies. Our analysis unveils
findings that are key for the design of detection and counter measurements approaches. 
}

\section{Introduction}

Online Social Networks (OSNs) have become popular communication platforms where people post messages about everything, from day-to-day conversations to their opinions about
noteworthy events. The size of the active user bases and the volume of data created daily on OSNs are massive.  Twitter, a popular micro-blogging site, has more
than 200 million active users, who post more than 500 million tweets a day~\cite{Protalinski:2013:Online}.  Notably, about 90\% of the Twitter users make their content public~\cite{cha:2010},
allowing researchers and companies to gather and analyze data at scale.  

The massive amount of public data offered by Twitter, associated with its simple and easy-to-use public API, have allowed the emergence of a new wave of applications that explore
real-time Twitter data to offer data mining and knowledge extraction services.  Indeed, in the last few years, there was a dramatic proliferation of applications and studies that monitor memes,
brands, people, products, and noteworthy events in Twitter, including political elections~\cite{Tumasjan:2010}, stock marketing fluctuations~\cite{Zhang:2011}, disease epidemic
outbreaks~\cite{Gomide:2011}, 
and natural disasters~\cite{Sakaki:2010}.

Although appealing as mechanisms to summarize information from the crowds, these services may be vulnerable to attacks that attempt to tamper their
statistics, by disseminating  misinformation in the network~\cite{Castillo:2011:ICT:1963405.1963500}.  For instance, consider that users of a Twitter-based service might be
interested in knowing what others think about a certain political candidate to formulate their own opinion.  In this scenario, one could try to create fake accounts which post tweets to dishonestly improve or damage the
public perception about this person, as an attempt to manipulate public opinion.  Recent efforts estimate that more than 20 million Twitter accounts were fake in
2013~\cite{20M-fake-users-Twitter} and even Twitter admitted that 5\% of its users are fake~\cite{Twitter-5percent-users-fake}.  More alarmingly, 
socialbots -- OSN accounts that are automatically controlled but are made
to look like real users -- are already being used to automate this kind of attack~\cite{Messias:2013,Harris:2013:Online,Boshmaf:2011}.

If socialbots could be created in large numbers, they can potentially
be used to bias public opinion, for example, by writing large amounts of fake messages and dishonestly improve or damage the
public perception about a topic.  There are already evidences of the use of socialbots to create a impression that emerging political movements are popular and
spontaneous~\cite{Ratkiewicz:2011}.  
Socialbots have also been used by political candidates during election campaigns to try to change the ``trending
topics''~\cite{Orcutt:2012:Online}, or to artificially increase their number of followers~\cite{Calzolari:2012:Online}. 
This scenario only gets worse when we consider the existence of socialbot sale services.\footnote{http://www.jetbots.com/}

Despite the large number of applications and services that rely 
on Twitter data and the emerging 
black markets that attempt to manipulate such data~\cite{Thomas:2013}, 
there are many unanswered questions related to 
socialbots infiltration in Twitter. 
For instance, \textit{Can socialbots really infiltrate Twitter easily?}, \textit{What are the characteristics of socialbots that
would enable them to evade current Twitter defenses?},
\textit{What strategies could be more effective to gain followers and influence?}, \textit{What automatic posting patterns could
be deployed by socialbots without being detected?},
and so on.

To answer these and other questions related to socialbots,
this study attempts to reverse engineer socialbot infiltration 
 strategies in Twitter.  Our methodology consists of
creating 120 socialbot accounts with different characteristics and infiltration strategies (e.g., gender specified in the profile,  
how active they are in interacting with users, 
the method used to generate their tweets), 
and investigating the extent to which these 
bots are able to infiltrate the Twitter social network
over the duration of a month.  
More specifically, we analyze which socialbot strategy is more successful 
in acquiring followers and provoking interactions 
(such as retweets and mentions) from other Twitter users.\footnote{Note that
our only objective in analyzing the factors that can enable
socialbot infiltration, is to improve defense mechanisms in OSNs.
We ensured that none of our socialbots actually posted any malicious
content in Twitter. Furthermore, all the socialbot accounts were deleted after
one month of experimentation.}
Then, we perform a $2^k$
factorial design experiment~\cite{jain} to quantitatively measure the extent to which each social strategy affect different infiltration metrics.

Our findings raise a huge alert about the vulnerability of many existing Twitter-based services.  
First, we show that out of the 120 socialbots we created, only 31\% 
could be detected by Twitter after a
period of one month of executing only automated behavior, 
suggesting that automated strategies are able 
to evade Twitter defense mechanisms.  
Second, we show that even socialbots
employing simple automated mechanisms successfully acquired hordes of followers 
and triggered a large number of interactions from other
users, making several bots to become relatively highly
influential according to metrics like Klout score~\cite{klout}. 
Therefore, our proposed approach to 
measure which bot strategy works better for infiltrating Twitter, may
be extremely valuable for the design of future defense mechanisms. 
Particularly, we found that higher Twitter activity 
(such as following users and posting messages) showed to be the
most important factor towards successful infiltration when bots 
target a random group of users.  
Other factors, such as the gender and the profile picture, may gain importance when
the attack is concentrated on a particular type of users.  

As a final contribution, we plan to {\it make our dataset available to the research community} by the time of publication of this article.  The dataset consists of the timeline of
activities and performance of infiltration of each of the 120 socialbots during the 30 days of experimentation.  To the best of our knowledge, this dataset is the first of its
kind, and would allow researchers to explore new aspects of socialbots infiltration in Twitter.

The rest of the paper is organized as follows. 
The next section briefly surveys related efforts.  
In Section 3, we present the methodology used to create the socialbots.  
Section 4 presents the metrics used to evaluate the infiltration performance of socialbots.
Section 5 describes which socialbots were able
to evade Twitter defenses, whereas Section 6 evaluates the socialbots 
configurations that achieve the best infiltration performance. 
Then, Section 7 describes a $2^k$ factorial
design experiment to quantitatively assess the relative importance of 
various
features in socialbot infiltration strategies.  
Finally, Section 8 discusses the implications of our findings
to future defense mechanisms and
directions of future work.

\section{Related work}

Different forms of spam have been observed in online systems~\cite{Heymann_fightspam_2007} and OSNs such as Twitter.  Specifically, there has been lot of attention on spam
in Twitter, which has been observed to be much more potent than conventional modes of spam such as email-spam~\cite{Grier:2010}.  In fact, several different types of spam activity
have been observed in Twitter, including spamming trending topics~\cite{benevenuto@ceas10}, polluters~\cite{Lee:2011}, link farming~\cite{ghosh:2012}, 
phishing~\cite{Benevenuto:2011}, content credibility issues~\cite{Castillo:2011:ICT:1963405.1963500}, and automated and fraudulent accounts~\cite{Zhang:2011,Thomas:2013}. 
Given that the present study focuses on socialbots in Twitter, 
the rest of this section is devoted to review studies related to socialbots.

~\\
\noindent\textbf{Bots in OSNs:} A particular form of OSN spam consists
of the spammers creating socialbots (or simply bots) 
which attempt to acquire influence and trust
in the OSN before engaging in malicious activities
such as spreading misinformation and manipulating public opinion.
There have been several attempts towards large-scale creation of bots
in OSNs, such as the Realboy project~\cite{realboy:Online} or the
Web Ecology project~\cite{web-ecology}.
Messias {\it et al.}~\cite{Messias:2013} created
bots capable of interacting with users on Twitter, and 
achieved significant scores according to
influence metrics such as Klout and
Twitalyzer.\footnote{http://twitalyzer.com}
Boshmaf {\it et al.}~\cite{Boshmaf:2011} designed a
social network of bots in order to conduct a large-scale
infiltration; the study demonstrated that OSNs
can be infiltrated with a success rate of up to 80\%.
In general, these efforts
demonstrate the vulnerability of Twitter to the infiltration of bots.

There have also been attempts for detection of bots in OSNs.
Chu {\it et al.}~\cite{Chu:2012} used machine
learning techniques to identify three types of accounts on Twitter
$-$ users, bots and cyborgs (users assisted by bots). 
They showed that the regularity of posting, the fraction of tweets with URLs and the posting medium used (e.g., external apps), provide evidence for the type of the account. 
Complementary to the detection of bots, 
Wagner {\it et al.}~\cite{Wagner:2012} created a machine learning
model to predict user's susceptibility to bot attacks 
using three different set of attributes of the user (network, behavior and
linguistic characteristics). Their results indicate that
users more ``open'' to social interactions are more susceptible to attacks.
Subsequently, a similar study by Wald {\it et al.}~\cite {Wald:2013} 
found that the Klout score, number of followers and friends, 
are good predictors of whether a user will interact with bots.

~\\
\noindent\textbf{Two perspectives of studying spam:}
The numerous studies on spam in online forums can be broadly
divided into two classes based on the perspective from 
which the study is conducted. 
A large majority of the studies (including the ones
stated above) are from the {\it perspective of 
those who build spam defense mechanisms}, such as 
developing classifiers for spammers and bots~\cite{Chu:2012,Lee:2011}.
However, there have been a few studies which were conducted
from the {\it perspective of spammers}; these studies
essentially attempt to reverse engineer the strategies
of spammers in order to gain insights which can help
to develop better spam defenses.

Most of such studies (from the perspective of spammers) 
have been on email spam, and spam in the Web. 
Pitsillidis {\it et al.}~\cite{Pitsillidis:2010} attempted to filter
spam emails by exploiting the perspective of the spammers $-$
they instantiated botnet hosts in a controlled environment, and 
monitored spam emails as they were created, and thus inferred 
the underlying template used to generate such emails.
Stone-Gross {\it et al.}\cite{Stone-Gross-botmaster} studied
a large-scale botnet from the perspective of the botmaster,
and analyzed the methodologies used in orchestrating spam
email campaigns.
Gyongyi \textit{et al.} studied link farms on the Web, which are
 groups of interconnected web-pages which attempt to boost
the rankings of particular web-pages. Specifically,
they investigated how multiple web-pages 
can be interconnected to optimize rankings~\cite{Gyongyi_linkspamalliances_vldb}. 

Almost all studies on Twitter spam has focused on understanding 
the dynamics of different forms of spam in Twitter, or on designing 
spam defense mechanisms.
To the best of our knowledge, there is no previous study
attempting to analyze the strategies of bots/spammers
from the perspective of the spammers themselves.
This is the motivation of the present study $-$
to reverse engineer socialbot strategies in the Twitter OSN.
We believe that this is complementary to all the aforementioned 
studies on Twitter spam, can
offer a novel perspective to building more effective
defense mechanisms against spam and bot accounts in the future.

\section{Methodology}\label{sec:methodology}


This study aims to reverse-engineer socialbot strategies in Twitter, and analyze how various characteristics of the socialbots impact their infiltration performance.
For this, it is necessary to 
create a set of socialbots in Twitter, which would attempt to infiltrate the network, and then
observe their behavior and infiltration performance.
This section details the steps followed in this infiltration experiment.

\subsection{Creation of socialbots} \label{sub:socialbot-creation}

For the experiments in this study, a set of 120 socialbots 
were created on Twitter.
The socialbots were implemented using the open-source 
Realboy project~\cite{realboy:Online}
which is an experimental effort to create `believable' Twitter bots
using simple automated techniques to follow other Twitter users
and post tweets (see~\cite{realboy:Online} for details).
Our 120 bots were created over a period of 20 days,
using 12 distinct IP addresses (10 bots were
operated from each IP address). 
Subsequently, starting from 10 days after the creation of the last bot, 
we monitored their behavior and
their interactions with other Twitter users over a period of 30 days.

~\\
\noindent{\bf Profile settings of socialbots:}
To make bot accounts look similar to real users in Twitter, we 
took the following steps while creating the bot accounts.
Each socialbot has a customized profile, which includes
a name, a biography, a profile picture, and a background.
The gender of the bot is also set to `male' or `female' using a proper name and profile picture.
Further, to ensure that when other Twitter users see our bot accounts,
they do not see a totally `empty' profile, the socialbots
are initially set to have a few followers, followings and tweets.
As detailed later in this section, 
the 120 socialbots are divided into groups based on the set of target
users they are assigned to follow.
Each bot initially follows a small number (randomly selected
between one and seven) of the most popular users
among the target users assigned to it.
Also, all socialbots assigned to the same target-set follow
each other, so that every bot account has some followers to start with.
Finally, every socialbot posted 10 tweets before attempting
to interact with other users.

~\\
\noindent{\bf Activity settings of socialbots:}
Since the objective of our socialbots 
is to infiltrate the social network, it is necessary 
that they interact with other users in the social network. 
For this, our socialbots can perform a
set of basic actions: 
(i)~post tweets, (ii)~retweet tweets posted by the users they follow, and
(iii)~follow users on Twitter. 
  
Specifically, a socialbot becomes `active' at pre-defined
instants of time; the gap between two such instants of activity
is chosen randomly (as detailed later in this section).
Once a socialbot becomes active, it performs the following two
actions -- 
(i)~with equal probability, the socialbot either posts a new tweet, 
or retweets a tweet that it has received from its followings, and 
(ii)~the socialbot follows a random number (between 1 and 5) of
the target users assigned to it, and follows some of the
users who have followed it  (if any)
since the last instant of activity.

Note that we attempt to ensure that our bots do {\it not} 
link to spammers or other fake accounts, 
which could make Twitter's spam defense mechanisms suspicious,
and potentially lead to suspension of the bot accounts.
For this, our bots only follow users from their respective
target-set, and some selected users from among
those who have followed them.
Since it is known that spammers in 
Twitter usually have far less number of followers
than the number of followings~\cite{benevenuto@ceas10,Lee:2011}, 
our socialbots follow back non-targeted users
only if 
those users have their number of followers 
greater than half the number of their followings.


\subsection{Attributes of the socialbots}  \label{sub:socialbot-attributes}


There are a number of attributes of a Twitter user-account which could
potentially influence how it is viewed
by other users.
Since analyzing the impact of all possible attributes
involves a high cost, we decided to focus on 
the following four specific attributes of the socialbot accounts:
(i) the gender mentioned in the bot's profile,
(ii) the activity level, i.e., how active the bot is in following users and posting tweets,
(iii) the strategy used by the socialbot to generate tweets, and
(iv) the target set of users whom the socialbot links with.

We set the socialbot accounts such that
they have diverse characteristics with respect to each of the four
attributes stated above, and then attempt to measure
whether any of these attributes can make a bot
more successful in terms of infiltration.
The rest of this section describes these attributes,  
and their distribution in the 120 socialbots created.

\subsubsection{Gender}

Out of the 120 socialbots created for our experiments,
half are specified to be male, while the other half are specified
to be female. Setting the gender of a socialbot involves using an appropriate name and profile picture.

\subsubsection{Activity level}
Here we aim at investigating whether more active bots 
are more likely to be successful in infiltration tasks, than less active ones. 
Note that while more active bots are more likely to be visible to other users,
they are also more likely to be detected by Twitter defense mechanisms;
hence there is a trade-off in deciding the activity level of socialbots.
For simplicity, we create socialbots 
with only two levels of activity, based on the interval
between two consecutive instants when a bot becomes 
`active' and engages in following users and posting tweets
(as stated earlier in Section~\ref{sub:socialbot-creation}):\\

\noindent  \textbf{(1) High activity:}
For these socialbots, the intervals between two consecutive
instances of activity are chosen randomly between 1 and 60 minutes.\\ 
\noindent \textbf{(2) Low activity:}
For these socialbots, the intervals between two consecutive
instances of activity are chosen randomly between 1 and 120 minutes.\\ 

Half of our 120 socialbots are configured with high activity,
while the other 60 are configured with low activity.
Also, all socialbots `sleep' between 22:00 and 09:00 Pacific time zone,
simulating the expected downtime of human users.

 
\subsubsection{Tweet generating strategy}

Making a socialbot to look like a 
legitimate user requires {\it automated} methodologies for generating well-written tweets
with relevant content.
We employ two different methodologies for generating tweets
from our bot accounts:\\

\noindent \textbf{(1) Re-posting:}  
As the name indicates, this method consists of re-posting
tweets that were originally posted by another user, as if
they were one's own.
A socialbot employing this strategy simply re-posts tweets
(written by other Twitter users) drawn from the random sample
of the Twitter stream.\footnote{Twitter provides a 1\% random sample
 of the complete tweet stream for public use.} 

Note that a
very large fraction of posts in Twitter are
day-to-day 
conversation~\cite{wagner_topic_expertise,sampling-cikm}.
Hence blindly re-posting {\it any} random tweet 
may result in
posting mostly conversational tweets,
which would not seem interesting to the target users (whom
the bot intends to interact with).
Thus, we adopted the following approach to increase the odds that the
tweets re-posted by our bots have content relevant to the target users.
For a particular bot, we extracted the top 20 terms 
that are most frequently used by the target users of that bot
(after ignoring a common set of English stop-words).
Subsequently, a bot considers a tweet for re-posting
only if it contains at least one of these top 20 terms.

\begin{figure}[tb]
\centering
\includegraphics[scale=.35]{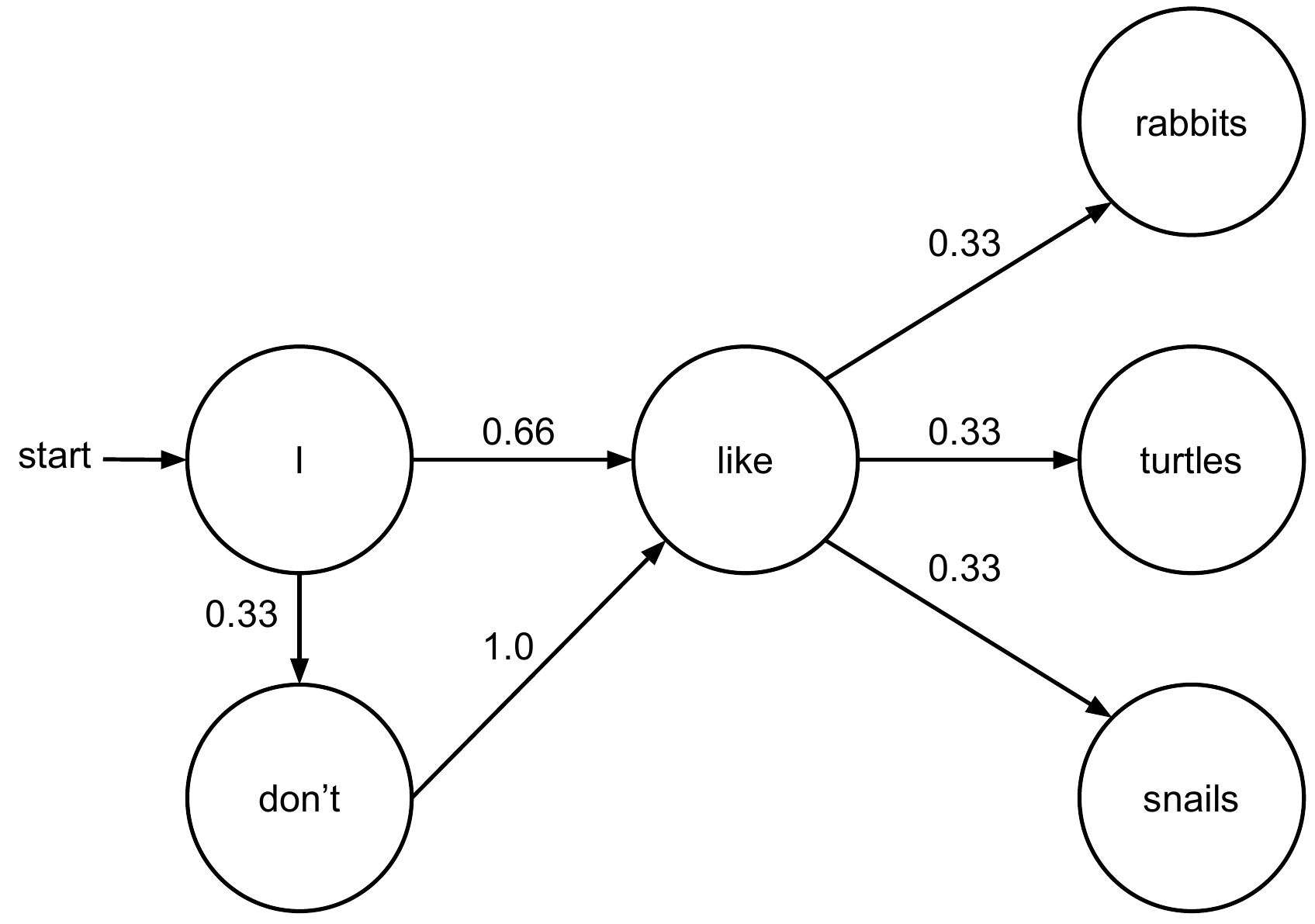}
\caption{Example of a bigram Markov chain -- to demonstrate the approach used to synthetically generate tweets posted by the socialbots.}
\vspace*{-2mm}
\label{fig:exemplomarkov}
\end{figure}

\noindent \textbf{(2) Generating synthetic tweets:}  
This approach synthetically generates
tweets that are likely to be as relevant as the tweets posted
by the target users.
Our approach uses a {\it Markov generator}~\cite{Barbieri:2012,Jurafsky:2000} $-$
a mathematical model used to generate text
that looks similar to the text contained
in a sample set of documents.
Figure~\ref{fig:exemplomarkov} shows an example of a bigram Markov generator, extracted from the sample set of documents \{``I like turtles'', ``I like
rabbits'' and ``I don't like snails''\}.
Here, the weight of an edge $w_i \rightarrow w_j$ denotes the 
probability that the word $w_j$ immediately follows word $w_i$, as
measured from the sample documents.\footnote{For instance, there is an edge
of weight $\frac{2}{3}$ between the nodes ``I'' and ``like'' since, out of 
the three occurrences of the word ``I'' in the sample documents, two occurrences
are immediately followed by the word ``like''.}
A possible text generated by the Markov generator in
Figure~\ref{fig:exemplomarkov} is ``I don't like rabbits''
(see~\cite{Barbieri:2012,Jurafsky:2000} for details of the method).

To increase the likelihood that the tweets generated by a socialbot are
considered relevant by the target users, we use a set of tweets 
recently posted by the target users of that socialbot, 
as the sample set to create the Markov generator.
Specifically, we use a trigram Markov generator, since
trigrams showed the best results when compared to $n$-grams of another order. 
We initially extract the empirical probability of occurrence of each
trigram in the sample set, then generate a Markov generator from the
obtained set of trigrams.

The main advantages of this method are that it 
does not require any human effort,
but it generates text containing the representative terms of the sample
set of documents. Thus, with a relatively high probability, the 
tweets generated by the socialbots 
are on the topics of interest of the target group.
However, the textual quality of the tweets may be low,
e.g., some tweets may be unfinished sentences.
Moreover, because of the way that the method has been
implemented it is unable to generate tweets containing 
user-mentions or URLs.
Table~\ref{table:markov-tweets} shows some example tweets generated by 
the Markov generator used in our experiment.\\


\begin{table}[tb]
\center
\small
\begin{tabular}{|p{0.9\columnwidth}|}
\hline 
I don't have an error in it :) \\ \hline
The amount of content being published this week :: the number of people who've finished this website but it makes it easier to argue that \\ \hline
Why isn't go in the morning! night y'all \\ \hline 
Night y'all ???! \\  \hline
take me to fernandos and you'll see \\  \hline
end aids now, the marilyn chambers memorial film festival I'd fix health care continues to outpace much of nation's issues move to the \\
\hline 
\end{tabular} 
\vspace*{-2mm}
\caption{{\bf Examples of tweets synthetically generated by the Markov generator}}
\label{table:markov-tweets}
\vspace*{-5mm}
\end{table}

\noindent Half of our 120 socialbots use only the reposting method, 
while the others utilize both the above methods, 
where each method has an equal probability to generate the next tweet.


\subsubsection{Target users}

Another factor which potentially affects the performance of socialbots
in infiltration tasks is the set of target users, i.e., 
the set of users with whom the socialbot attempts to interact and infiltrate. 
For instance, we wanted to check whether it is easier for socialbots
to infiltrate randomly selected users, or users who are similar
to each other in some way (e.g., users who are interested
in a common topic, or users who are socially connected among themselves).

As stated earlier, we wished to ensure that our socialbots do not
link to spammers or other fake accounts in Twitter.
Hence, we consider as a potential target user, only those user-accounts which
possess the following characteristics: 
(i)~are controlled by a human (as manually judged 
from the account's profile and the nature of the tweets posted), 
(ii)~post tweets in English, to ensure that they understand 
the language used by our bots and,
(iii)~are active, i.e., has posted at least one tweet since December 2013. 
We considered the following three groups of target users:\\

\noindent \textbf{Group 1:}  
Consisting of 200 users randomly selected
 from the Twitter stream random sample, and verified that they
 meet the above mentioned criteria.

\noindent \textbf{Group 2}:  
Consisting of 200 users who post tweets on a specific
topic. We decided to focus on a group of software developers; hence, 
we selected users from the Twitter random sample, who have posted at least one tweet 
containing any of the terms ``jQuery'', ``javascript'' or ``nodejs''. 
Subsequently, we randomly selected 200 accounts from among these users,
after verifying that they meet the criteria stated above.
Note that though we focus on software developers, 
the study could be conducted on groups of users  
interested in any arbitrary topic.

\noindent \textbf{Group 3}:  
Consisting of 200 users who
post tweets on a specific
topic, and are socially connected among themselves. 
As the topic, we again focus on software developers.
Here we started with a `seed user' -- {\it @jeresig}, who is an 
influential software developer on Twitter, and creator of `jQuery' --
and collected the 1-hop neighborhood of the seed user.
From among these users, we extracted 200 users whose profiles 
show that they are software developers, who satisfy
the criteria stated above, 
and whose social links form a 
dense sub-graph in the Twitter social network.\\ 

\begin{figure}[tb]
  \centering
  \subfloat[Age of the user account]{\includegraphics[width=.24\textwidth, height=3cm]{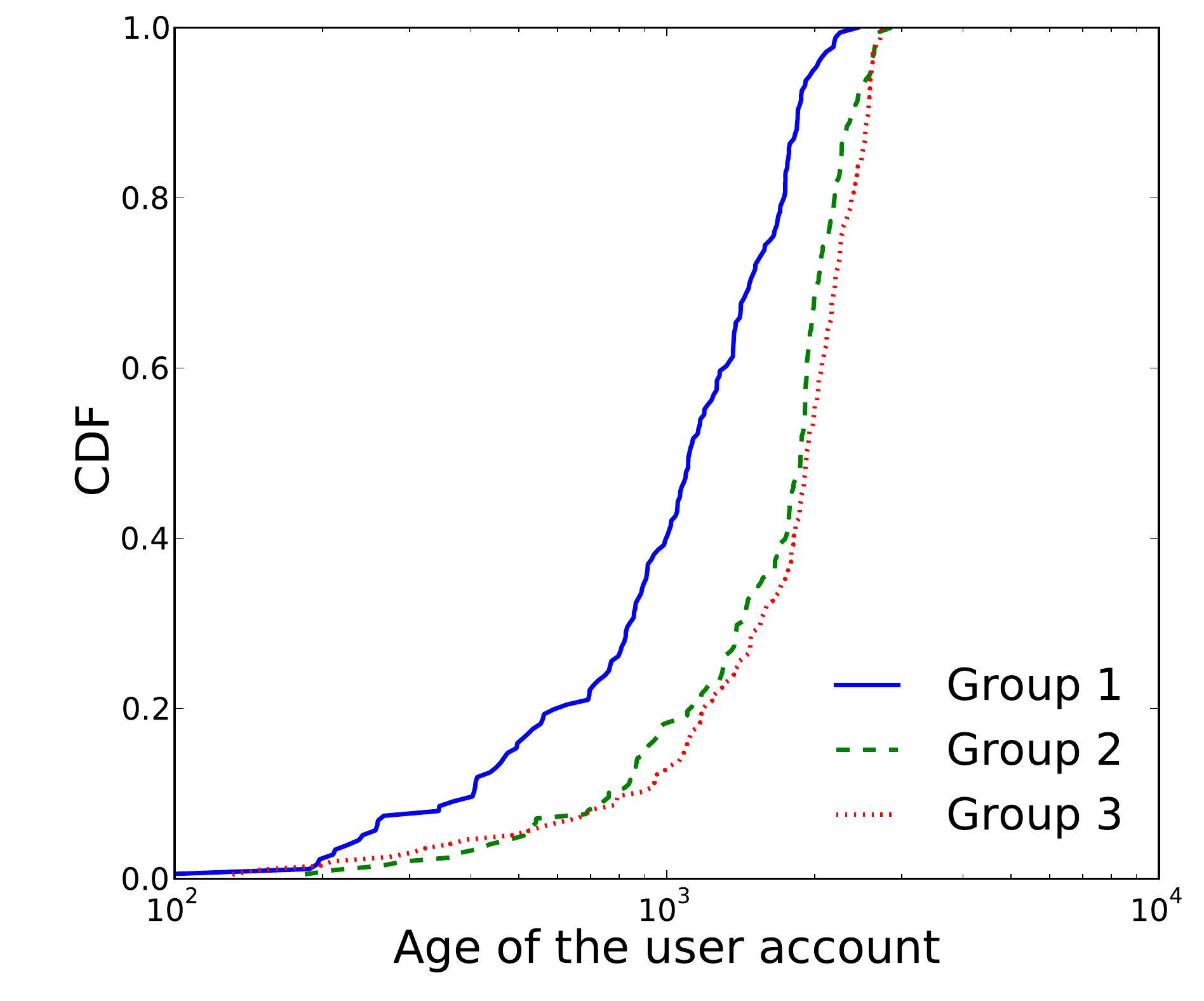}}
  \subfloat[Number of tweets posted]{\includegraphics[width=.24\textwidth, height=3cm]{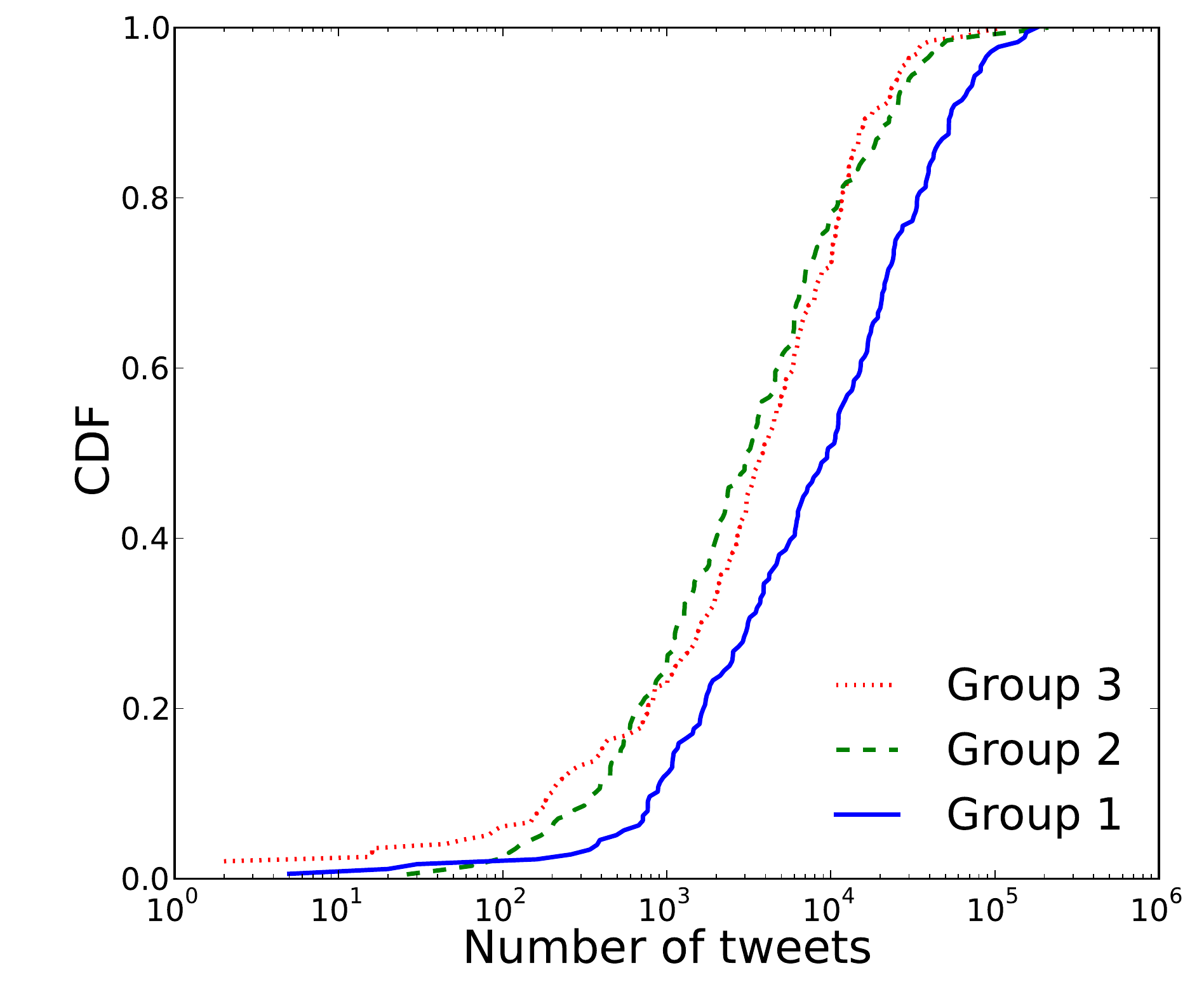}}
 
  \subfloat[Number of followers]{\includegraphics[width=.24\textwidth, height=3cm]{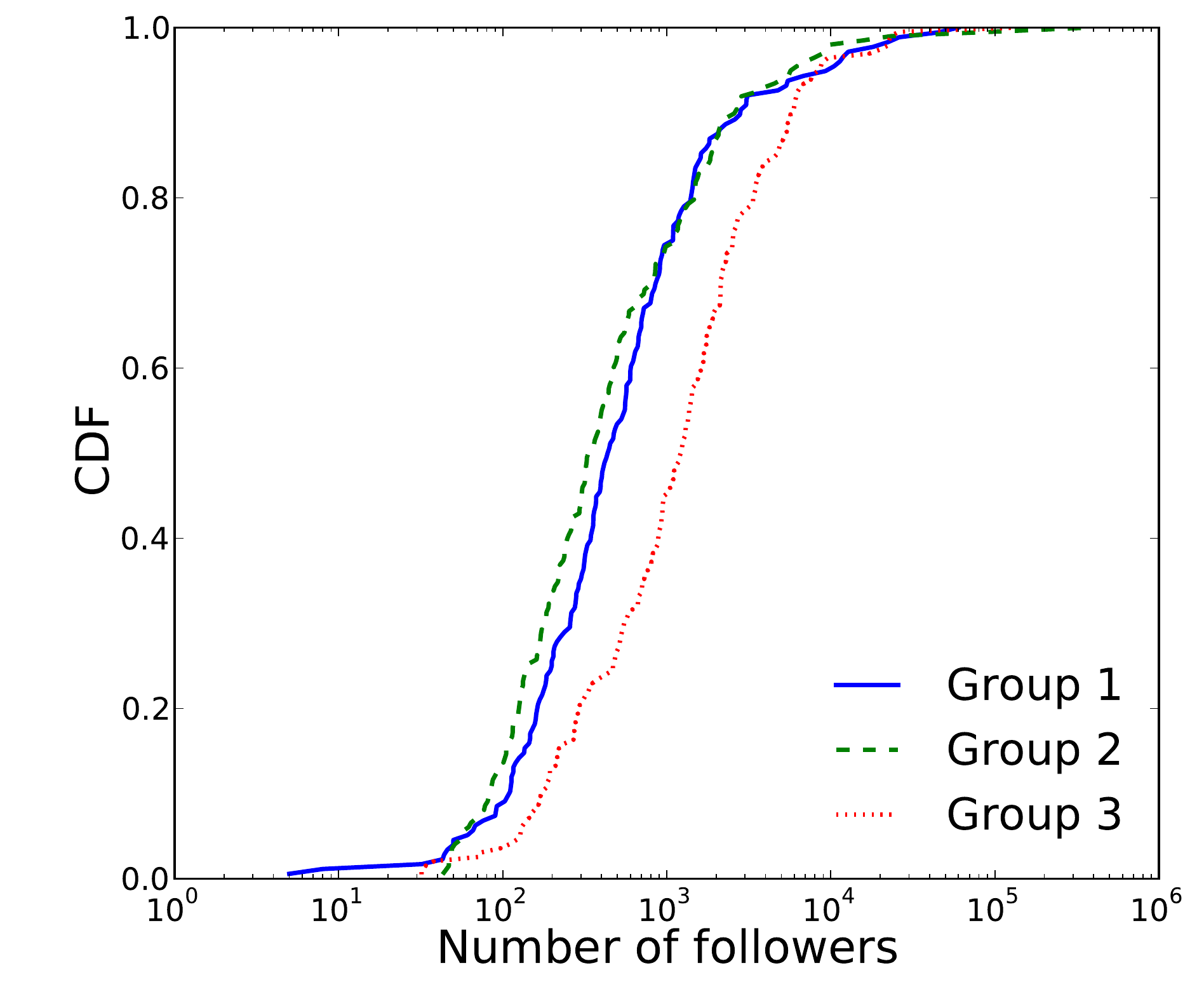}}
  \subfloat[Klout score]{\includegraphics[width=.24\textwidth, height=3cm]{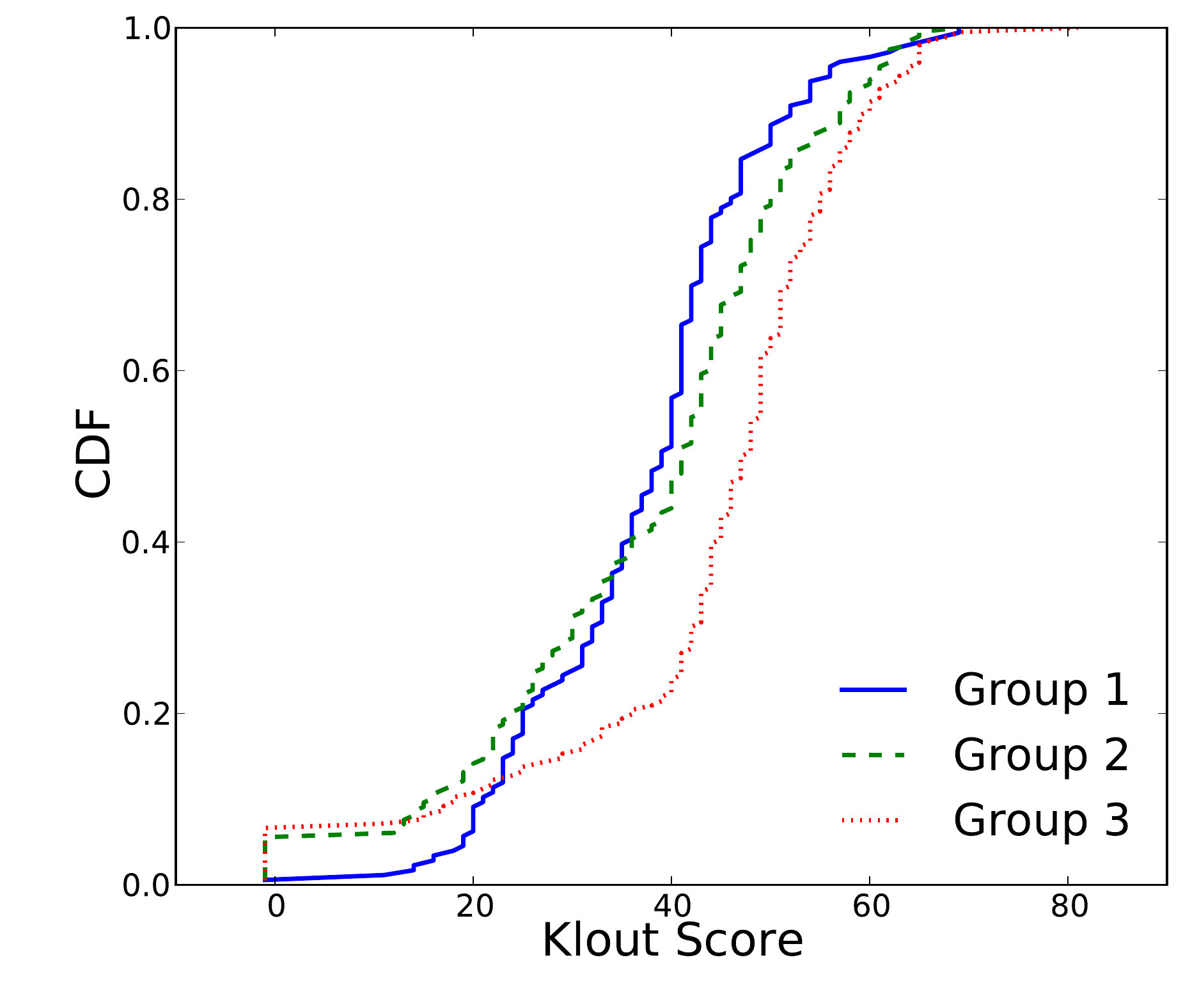}}
  \caption{Comparing the three groups of target users: CDFs for (i)~age of the user-accounts, (ii)~number of tweets posted by the users, (iii)~number of followers, and (iv)~Klout Score of the users in the three groups.}
  \label{fig:targetsattrs}
\vspace*{-5mm}
\end{figure}

\noindent Out of the 120 socialbots, 40 were assigned to 
each group of target users.
To bring out the differences among the three 
groups
(selected as described above), 
we conducted a brief characterization of each group.
Figure~\ref{fig:targetsattrs} shows distributions of the users in the three target groups
according to: (i)~the age of their accounts, 
(ii)~the total number of tweets posted during their life-time,
(iii)~their number of followers, and (iv)~their Klout scores.
We found that users in group 1 have relatively newer accounts than the other groups (Figure~\ref{fig:targetsattrs}(a));
however, they are more active in posting tweets (Figure~\ref{fig:targetsattrs}(b)).
Further, users in group 3 are more influential
than the other groups, i.e., have a greater number of 
followers (Figure~\ref{fig:targetsattrs}(c)) and 
higher Klout scores (Figure~\ref{fig:targetsattrs}(d)).

The justification behind our choices of target users is 
as follows. First, we intend to check whether it is easier for socialbots
to interact with and infiltrate {\it heterogeneous} groups of users 
(Group 1), or a particular category of users based on common 
interests (e.g., software developers, as in Group 2 and Group 3). 
Second, we wanted to compare the relative difficulty in 
infiltrating a group of users who are socially well-connected
among themselves, versus users who are not socially connected.
For this, we select Group 2 and Group 3 as stated above, so that
both groups of users are interested in the same topic, but 
Group 3 is densely connected in the social network while Group 2 is not.

\noindent Thus, the 120 socialbots created for this experiment
were configured with different strategies
for the four chosen attributes. 
The subsequent sections will analyze which of these
strategies result in better infiltration performance.

\section{Measuring Infiltration Performance}\label{sec:metrics}

The objective of this study is to investigate whether, and to what extent,
various socialbot strategies are able to infiltrate
the Twitter social network.
Naturally, we need some metrics to quantify the infiltration
performance of socialbots, so that the performance
of different strategies (employed by the socialbots) can be compared.
To quantify infiltration performance
we use the following three metrics, measured at the end
of the duration of the experiment.\\


\noindent {\bf (1) Followers acquired by the socialbot:} 
We count the number of followers acquired by the socialbot,
which is a standard metric for estimating the popularity /
influence of users in the Twitter social network~\cite{cha:2010}.

\noindent {\bf (2) Klout score acquired by the socialbot:}  
Klout score~\cite{klout}
is a popular metric for the online social influence of a user.
Though the exact algorithm for the metric is not known publicly,
the Klout score for a given user is known to consider various data points
from Twitter (and other OSNs, if available), 
such as the number of followers and followings of the user, 
retweets, membership of the user in Lists, how many spam / dead accounts 
are following the user, how influential are the people who retweet / mention
the user, and so on~\cite{klout-wiki}.
Klout scores range from 1 to 100, with higher scores implying 
a higher online social influence of a user.

\noindent {\bf (3) Message-based interactions with other users:}  
We measure the number of times other users interact with a socialbot
through messages (tweets) posted in the social network.
Considering the different types of 
message-based interactions allowed in Twitter, we specifically count the total
number of times some user @mentions the bot, or replies to the bot,
or retweets / favorites a tweet posted by the bot.
This metric estimates the {\it social engagement}
of the bot, which is defined as the extent to which a user
participates in a broad range of social roles and 
relationships~\cite{avison-mental-health}.\\

 

\noindent If a bot scores well in terms of the above metrics, 
it implies that the tweets posted by this bot are more likely
to be visible, e.g., more likely to be included in Twitter search results,
and is hence more likely to affect the opinion of other users
(which are common goals of bots in social networks).

The subsequent sections measure the success of various socialbot
strategies in infiltrating the social network according to
the metrics specified above.

\section{Can socialbots infiltrate Twitter?} \label{sec:success}

We first investigate whether, and to what extent, 
socialbots can infiltrate the Twitter social network. 
For a socialbot to successfully infiltrate
the network, it needs to achieve the following two objectives. 
(i)~evade detection by Twitter's spam defense mechanisms 
which regularly check for and suspend accounts exhibiting 
automated activities~\cite{twitter-shut-spammers}, and
(ii)~acquire a substantial level of popularity and influence
in the social network, and 
interact with large number of other users, i.e., achieve
high scores in the metrics described in Section~\ref{sec:metrics}\\
In this section, we investigate how socialbots
performed with respect to the above objectives.

\subsection{Socialbots can evade Twitter defenses}


We start by checking how many of the 120 socialbots could be detected
by Twitter. 
We find that over the 30 days during which the experiment
was carried out, 38 out of the 120 socialbots were 
suspended. This implies that although all our socialbots
actively posted tweets and followed other users during this period,
 as many as 69\% of the socialbots could {\it not}
be detected by Twitter spam defense mechanisms.

\begin{figure}[tb]
     \centering
     \includegraphics[width=\columnwidth, height = 4cm]{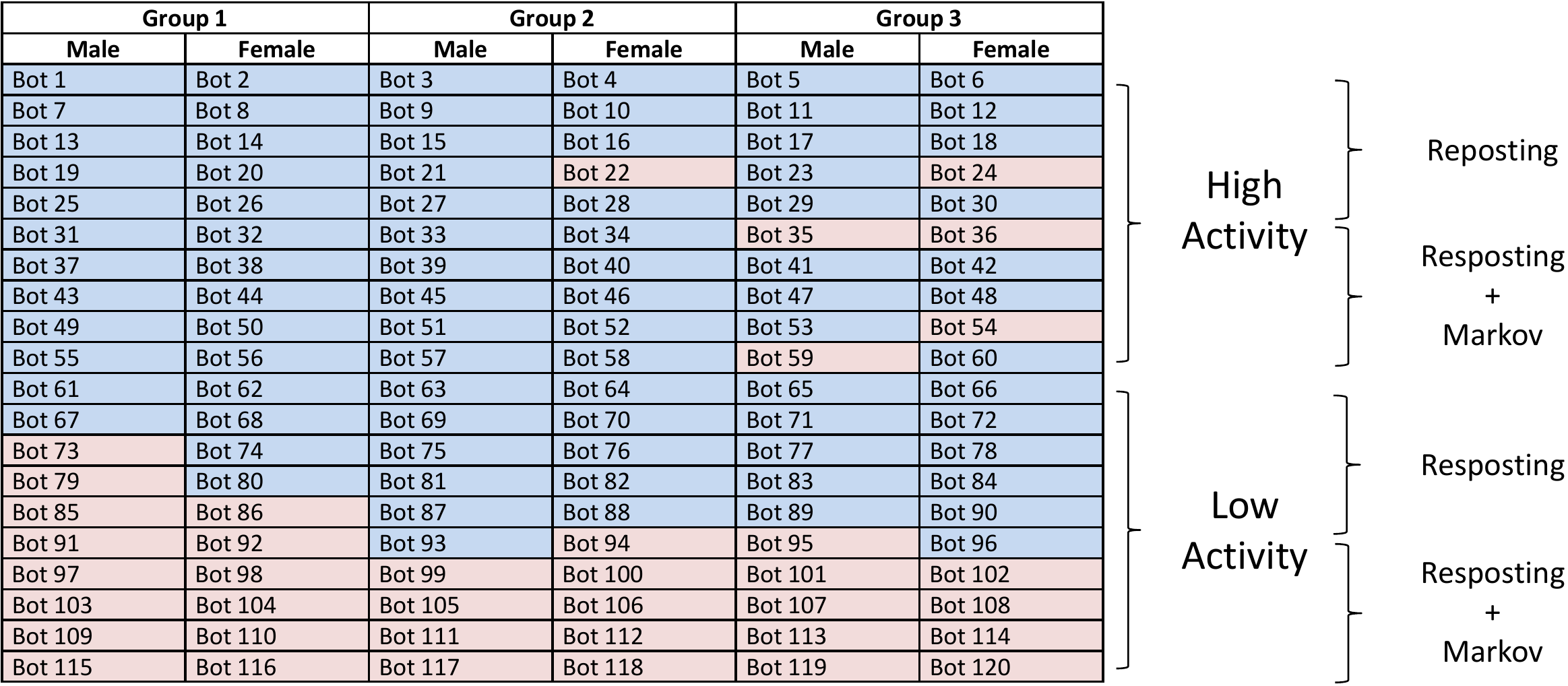}
     \caption{Distribution of attributes of the 120 socialbots created for the infiltration experiment, showing those socialbots which were detected and suspended by Twitter during the experiment (shown in red color). Note that 69\% of the socialbots (shown in blue color) could not be detected by Twitter.}
     \label{fig:suspended}
\end{figure}

We next analyze which of the 120 socialbots could be detected by
Twitter.
Figure~\ref{fig:suspended} shows the distribution of the four attributes
-- gender, level of activity, tweeting methodology, 
and target group of users followed -- among the 120 socialbots created. 
The socialbots are indicated by numeric identifiers in the same order in which 
they were created, i.e., Bot1 was created first and Bot120 was the last socialbot created.
The socialbots which were suspended by Twitter during the experiment (one month)
are indicated in red color, while the socialbots which could not be detected
by Twitter are shown in blue color. 

We note that a large majority of the socialbots which were suspended
were the ones which were {\it created at the end of the account creation
process} (with ids between 90 and 120).
This is probably because by the time these accounts were created,
Twitter's defense mechanisms had become
suspicious of several accounts being created from the 
same block of IP addresses.\footnote{As stated in Section~\ref{sec:methodology}, 
we used 12 distinct IP addresses to create the 120 socialbots, i.e., 10 accounts were operated from each IP address.}
We also find that the socialbots which used the 
Markov-based posting method were more likely to be suspended.
This is expected, since about half of the tweets posted by these
accounts were synthetically generated, and hence likely to be
of low textual quality.

However, the Twitter defense mechanisms could detect 
only a small fraction of the socialbots which 
were created early, and which adopted the re-posting strategy, i.e., 
re-posted others' tweets.
These statistics highlight that existing defense mechanisms
are of only limited use in detecting socialbots which employ 
simple but intelligent strategies for posting tweets
and linking to other users.
Further, the relatively low fraction of suspended accounts
justifies our strategies such as re-posting others' tweets, and
ensuring that our socialbots do not
link with large number of spammers and fake accounts 
(as described in Section~\ref{sec:methodology}).


\begin{figure*}
  \centering
  \hspace*{-5mm}
  \subfloat[{\bf Number of followers}]{\includegraphics[width=.3\textwidth, height=3cm]{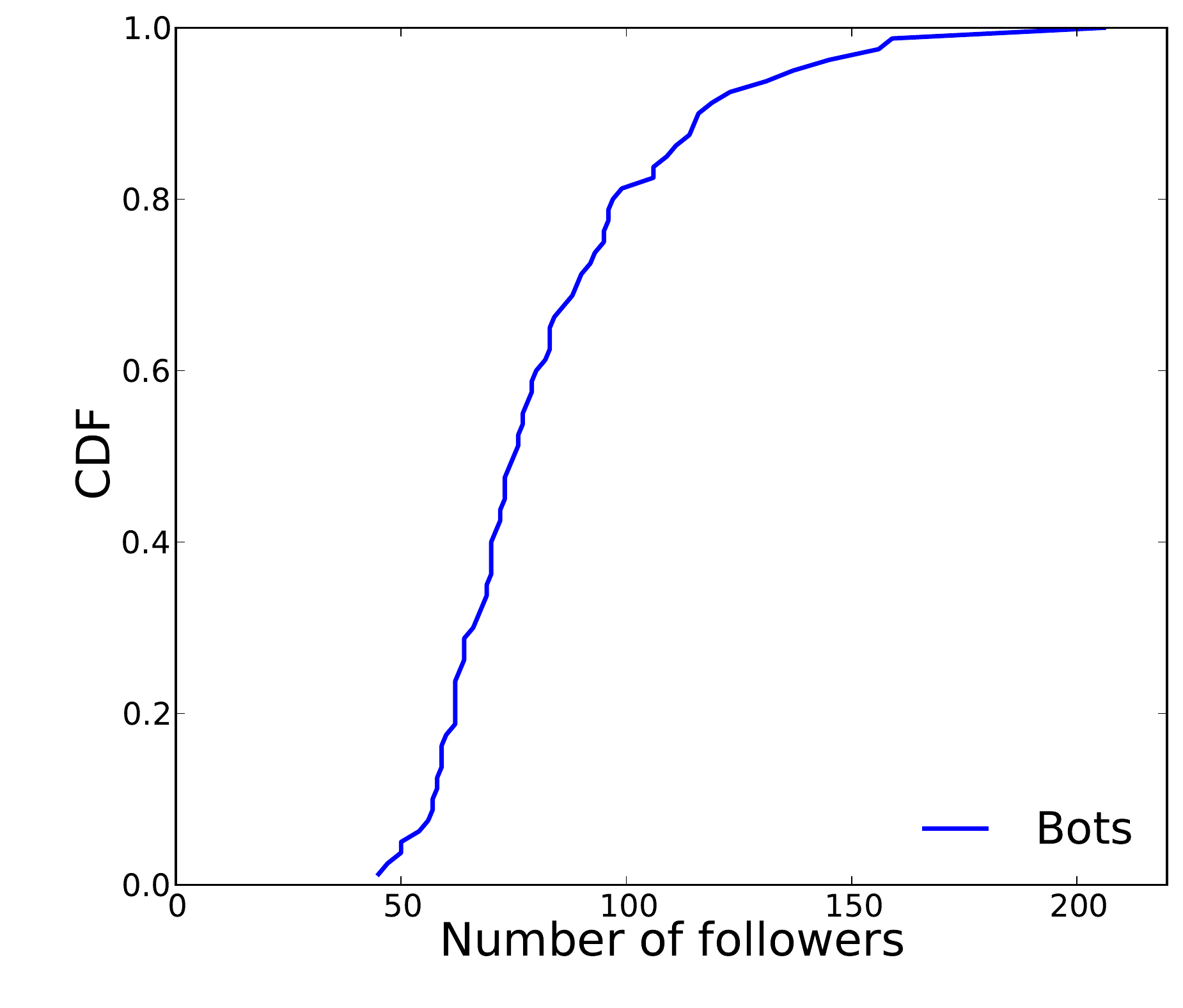}}
  \hfil
  \subfloat[{\bf Klout Score}]{\includegraphics[width=.3\textwidth, height=3cm]{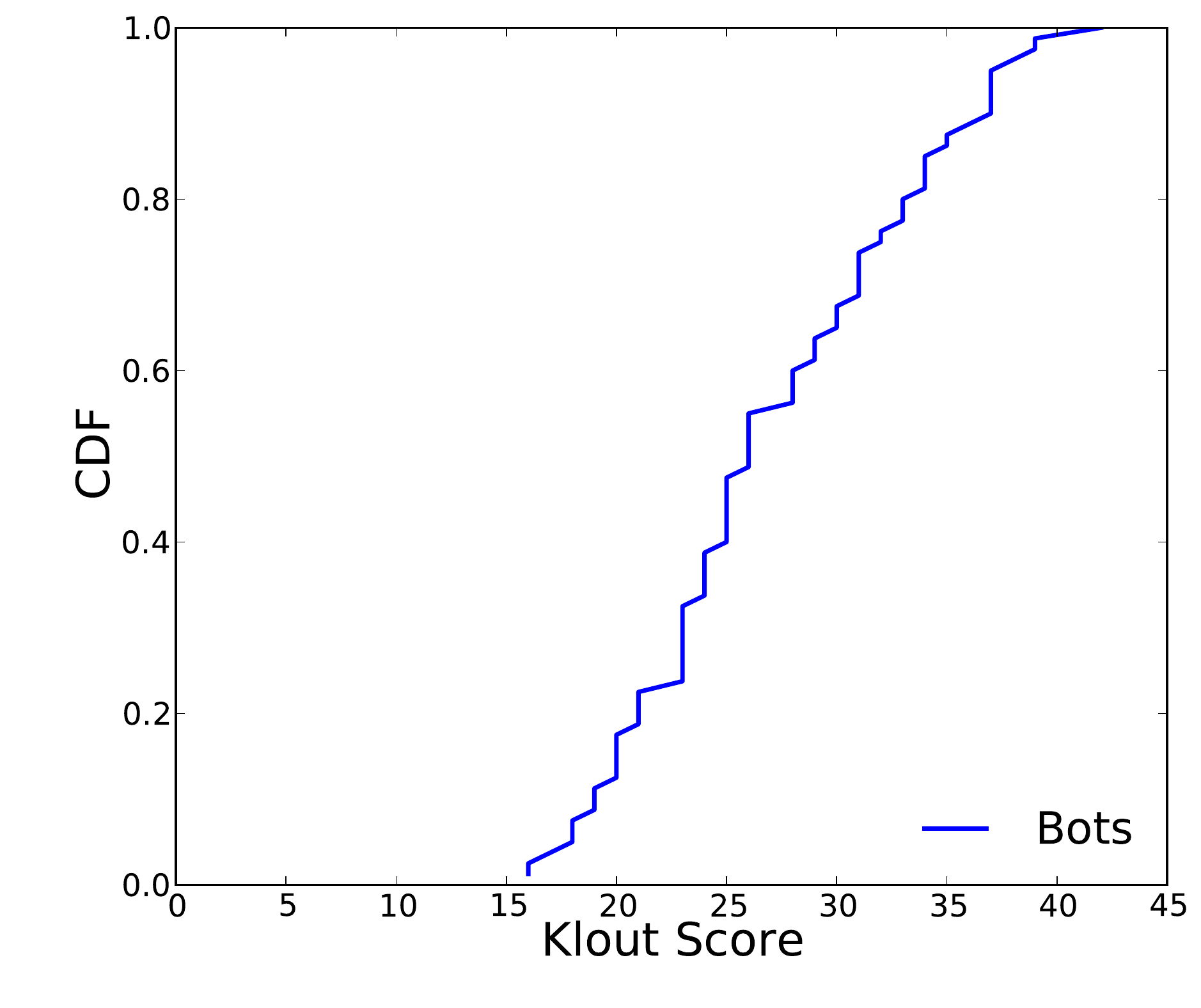}}
  \hfil
  \subfloat[{\bf Message interactions}]{\includegraphics[width=.3\textwidth, height=3cm]{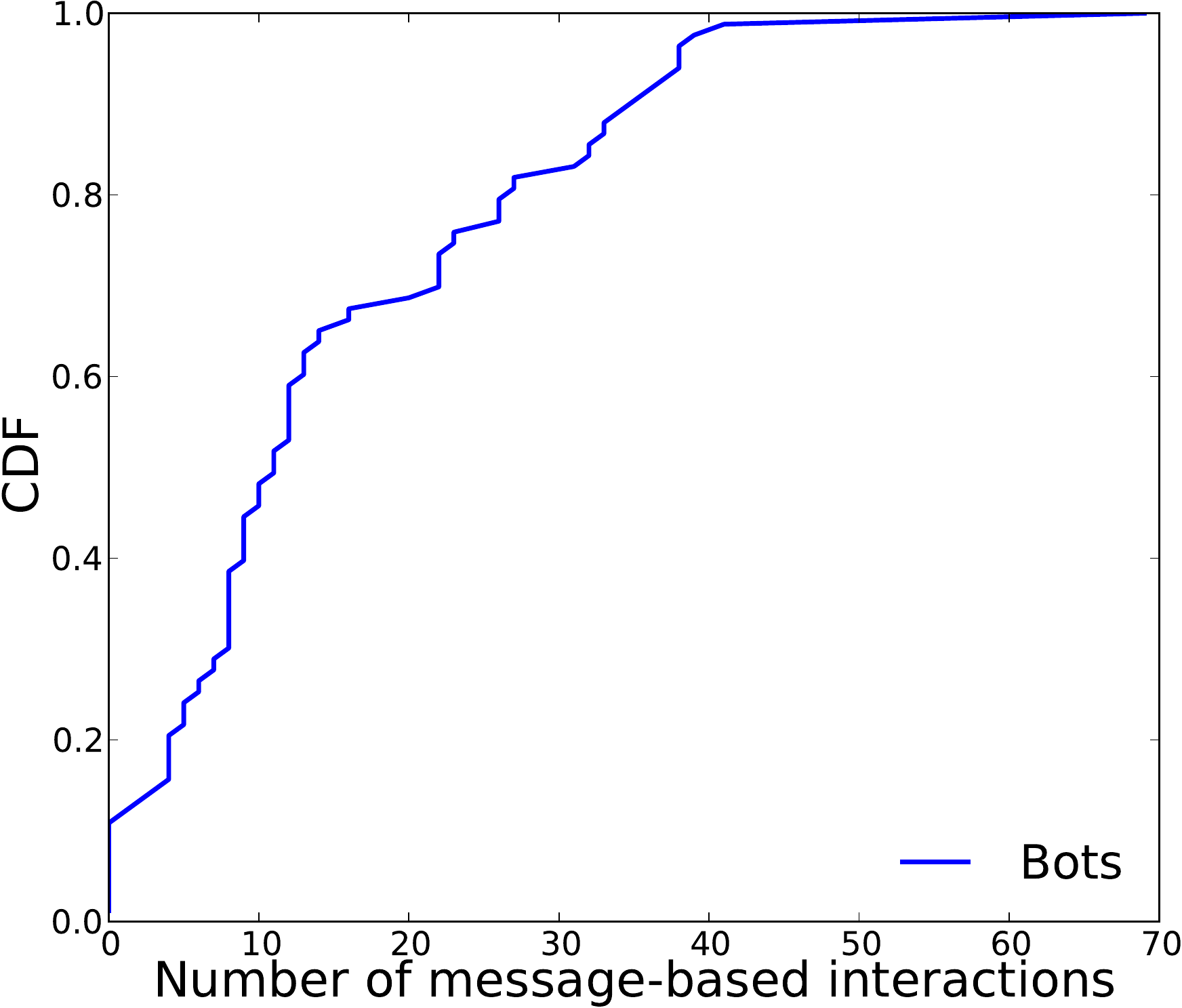}}
  \vspace*{-2mm}
  \caption{Infiltration performance of our socialbots: CDFs for (i) number of followers, (ii) Klout Score, and (iii) number of message-based interactions with other users.}
  \label{fig:bots-cdf}
\end{figure*}

\subsection{Bots can infiltrate Twitter successfully}

We next check to what extent socialbots can infiltrate
the Twitter social network, and whether they can gain
relatively high scores according to the metrics stated in Section~\ref{sec:metrics}.

Over the duration of the experiment, the 120 socialbots created by us 
received in total 4,999 follows from 1,952 distinct users,
and 2,128 message-based interactions from 1,187 distinct users.
Figure~\ref{fig:bots-cdf} shows the distribution of the 
number of followers, the Klout score and the number of message-based
interactions acquired by the socialbots at the end of the experiment.
It is evident that a significant fraction of the socialbots acquire
relatively high popularity and influence scores. 
Within just one month (the duration of the experiment), 
more than 20\% of the socialbots
acquired more than 100 followers (Figure~\ref{fig:bots-cdf}(a)); 
it can be noted that 46\% of the
users in Twitter have less than 100 followers~\cite{twitter-46pc-lt100followers}.
Furthermore, Figure~\ref{fig:bots-cdf}(b) shows that 20\% of
the socialbots acquired Klout scores higher than 35 within only one month.


Table~\ref{table:klout-comparison} compares the Klout scores acquired by the three socialbots that acquired the highest Klout scores\footnote{The three socialbots which acquired the highest Klout scores have common characteristics -- all of them had their gender specified as `female', were highly active, used only reposting as the mechanism for posting tweets, and followed Group 2 of target users (see Section~\ref{sec:methodology}).} 
with some real and active Twitter users -- members of the COSN
community. 
We find that the socialbots achieved Klout scores of the
same order of (or, at times, even higher than) several of these well-known academicians and social network researchers.  
Additionally, these socialbots also acquired higher Klout scores than the two bots developed in the prior study~\cite{Messias:2013}.
 
Note that the Klout scores of our socialbots were acquired over only one month (the duration of the experiment), whereas the real users in Table~\ref{table:klout-comparison} have
accumulated influence over several years.  
Besides, most of the real users in Table~\ref{table:klout-comparison} have accounts in multiple OSNs (e.g., Twitter, Facebook, LinkedIn),
all of which contribute towards their Klout scores; on the other hand, the Klout scores of our socialbots are only contributed by their Twitter accounts.  In spite of the above
limitations, the scores of our socialbots highlight the fact that 
bots employing relatively simple automated strategies 
to follow users and post tweets, can achieve
significantly high levels of popularity and social
engagement in Twitter network.

\begin{table}[htb]
\center
\small
\begin{tabular}{p{0.35\columnwidth}|p{0.45\columnwidth}|l}
\hline
User    & Description  & Klout \\
\hline
Carlos Castillo & COSN'14  TPC member & 56 \\		
Ben Zhao & COSN'14  TPC member  & 52 \\		
Ashish Goel  & COSN'14  TPC member  & 46  \\ 
Francesco Bonchi & COSN'14  TPC member  & 44  \\
Winter Mason & COSN'14  TPC member  & 44  \\
PK & COSN'14 Publicity Chair & 43  \\ 
Mirco  Musolesi & COSN'14 TPC member & 43  \\ 
{\bf Bot 28} & {\bf Socialbot in this study}   & {\bf 42}          \\
{\bf Bot 4}  & {\bf Socialbot in this study}  & {\bf 39}          \\
{\bf Bot 16} & {\bf Socialbot in this study}  & {\bf 39}          \\
Christo Wilson & COSN'14 Budget Chair & 38 \\  
{\it scarina} & {\it Bot developed in~\cite{Messias:2013}} & {\it 37.5}        \\
Fabricio Benevenuto & COSN'14  TPC member & 27      \\
Paolo Boldi  & COSN'14  TPC member  & 19      \\
{\it fepessoinha} & {\it Bot developed in~\cite{Messias:2013}} & {\it 12.3}        \\
\hline
\end{tabular}
\caption{Comparison of Klout scores of some of our socialbots with members of the COSN community, and bots developed in the prior study~\cite{Messias:2013}.} 
\label{table:klout-comparison}
\end{table}

\section{Evaluating Infiltration Stra\-tegies} \label{sec:performance}

The previous section showed that a significant fraction
of the socialbots are indeed able to infiltrate and gain popularity
in the Twitter social network.
This section analyzes which socialbot 
strategies lead to better infiltration performance.
Recall from Section~\ref{sec:methodology}
that the socialbots were configured with multiple strategies
for each of the four attributes -- 
gender, activity level, tweet-posting method, and type of target users.
We now investigate which strategies for each of the four attributes
yields the best infiltration performance.
Note that the results stated in this section (and the next)
consider those socialbots which were not suspended
by Twitter during the experiment (as described in Section~\ref{sec:success}).


\subsection{Gender}

\begin{figure*}[tb]
  \centering
  \hspace*{-5mm}
  \subfloat[{\bf Number of followers}]{\includegraphics[width=.33\textwidth, height=3.5cm]{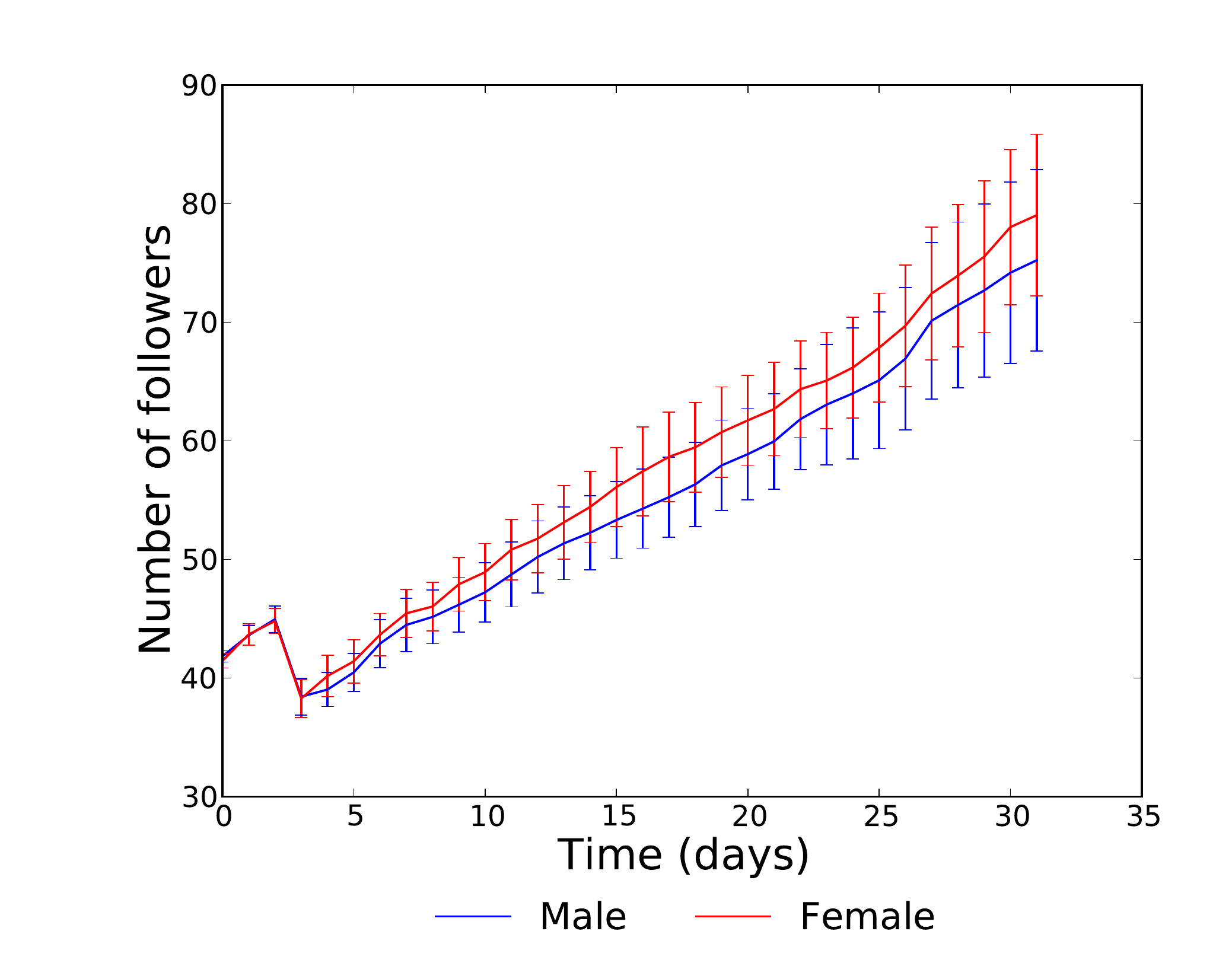}}
  \hfil
  \subfloat[{\bf Klout Score}]{\includegraphics[width=.33\textwidth, height=3.5cm]{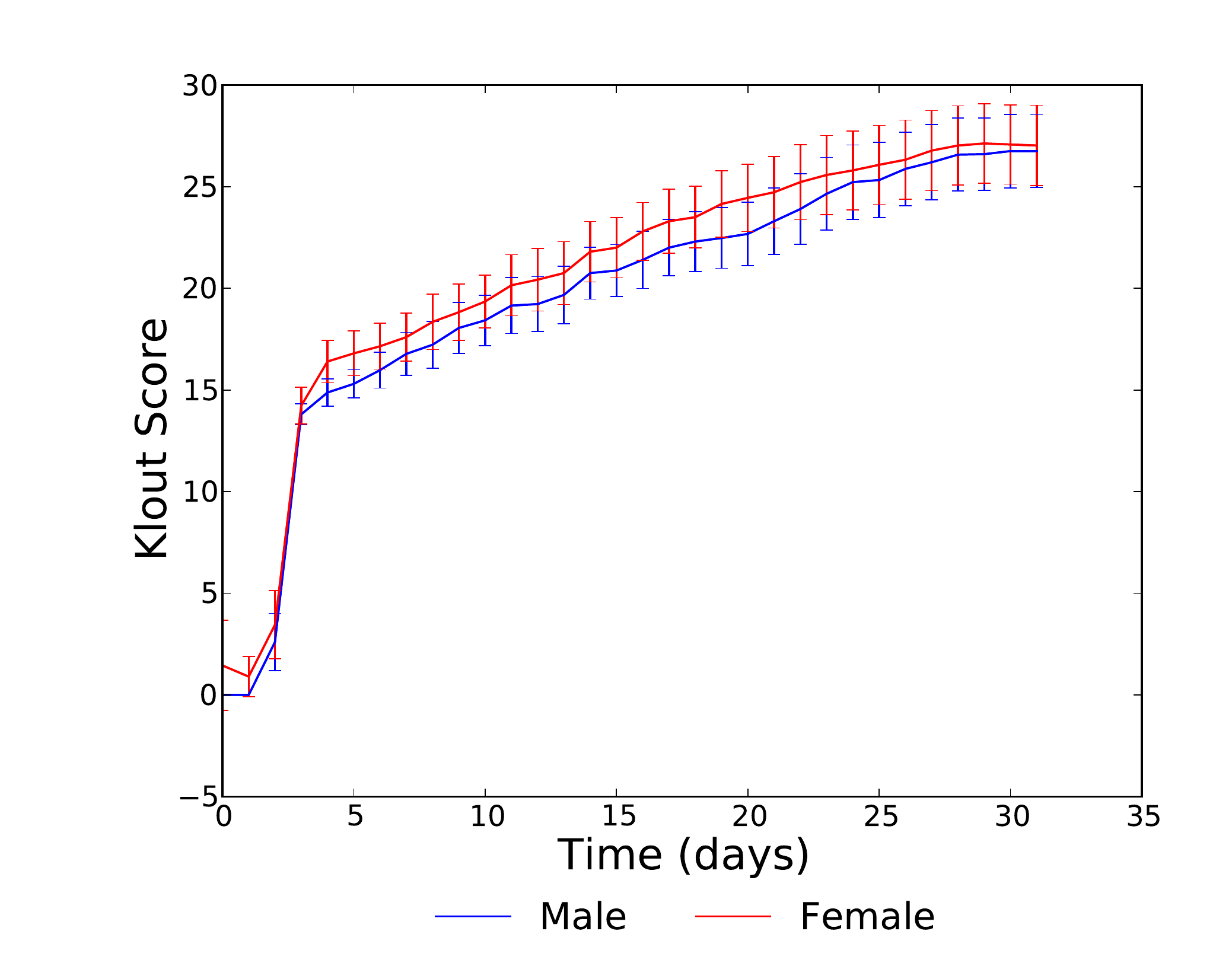}}
  \hfil
  \subfloat[{\bf Message interactions}]{\includegraphics[width=.33\textwidth, height=3.5cm]{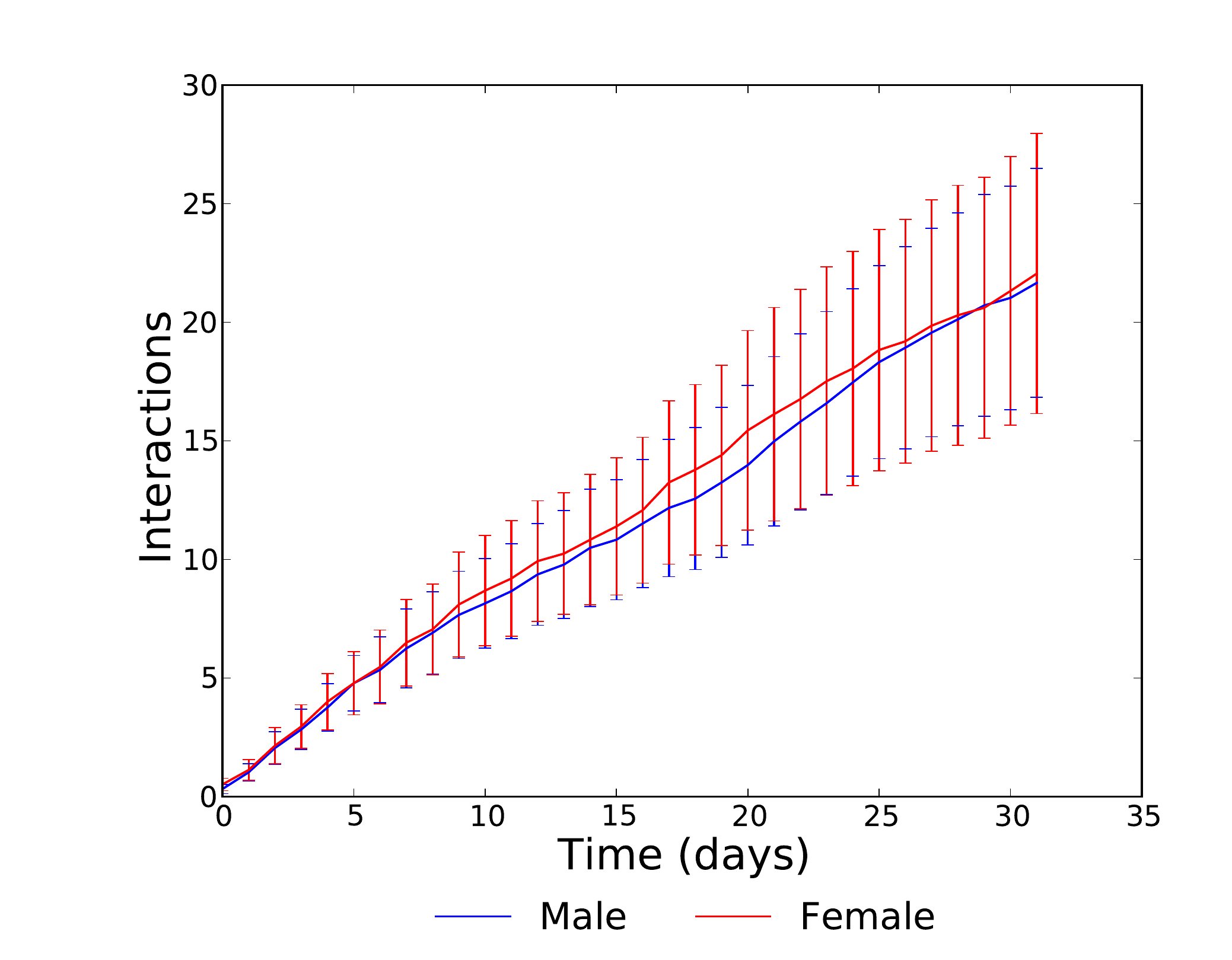}}
\vspace*{-2mm}
  \caption{Infiltration performance of socialbots of different genders through the duration of the experiment: (i)~mean number of followers acquired, (ii)~mean Klout score acquired, and (iii)~mean number of message-based interactions with other users. The curves represent the mean values, which the error bars indicated the 95\% confidence intervals.}
  \label{fig:performance-gender}
\vspace*{-5mm}
\end{figure*}

We start by analyzing the impact of the gender of the socialbots in our experiments. Figure~\ref{fig:performance-gender}(a) and (b) respectively show
the mean number of followers and the Klout score acquired by
the male and female socialbots over each day during our experiment.
In these figures, the curves represent the mean values considering all 
the socialbots of a particular gender (on a given day during the experiment),
and the error bars indicate the 95\% confidence intervals of the mean values.
We find that there is no significant difference 
in the popularity acquired by socialbots of different genders.



We next turn to the message-based interactions of the male and female
socialbots with other users.
Figure~\ref{fig:performance-gender}(c) 
shows the mean number of interactions of the socialbots 
on each day during the experiment. 
Again, we observe that users interacted almost equally
with socialbots of both genders.\footnote{The number of 
distinct users who interacted 
with the female socialbots (1,697),
was, in fact, slightly higher than the number who interacted
with the male socialbots (1,528).
But, as evident from  Figure~\ref{fig:performance-gender}(c),
this difference is not a significant one.}


The above results imply that the gender specified in the account profile 
does {\it not} significantly influence the infiltration performance of the socialbots. 
It can be noted that, in this section, we are considering
all the socialbots (and interactions with all target users) together.
Later in Section~\ref{sec:attributes}, when we separately analyze the performance of
socialbots in infiltrating each group of target users, 
we shall see that the gender attribute is indeed significant 
for some specific target groups.





\subsection{Activity Level}

\begin{figure*}[tb]
  \centering
  \hspace*{-5mm}
  \subfloat[{\bf Number of followers}]{\includegraphics[width=.33\textwidth, height=3.5cm]{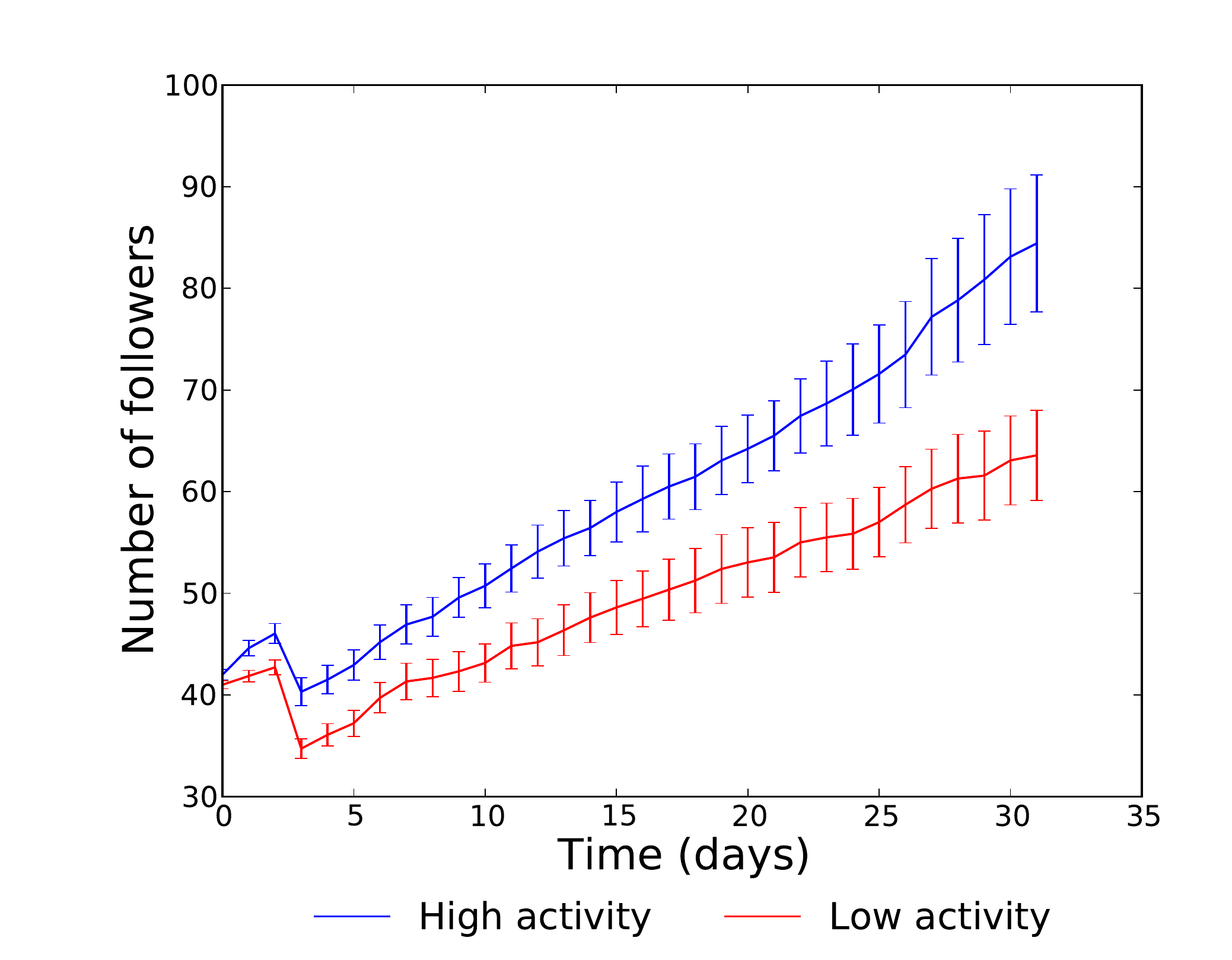}}
  \subfloat[{\bf Klout Score}]{\includegraphics[width=.33\textwidth, height=3.5cm]{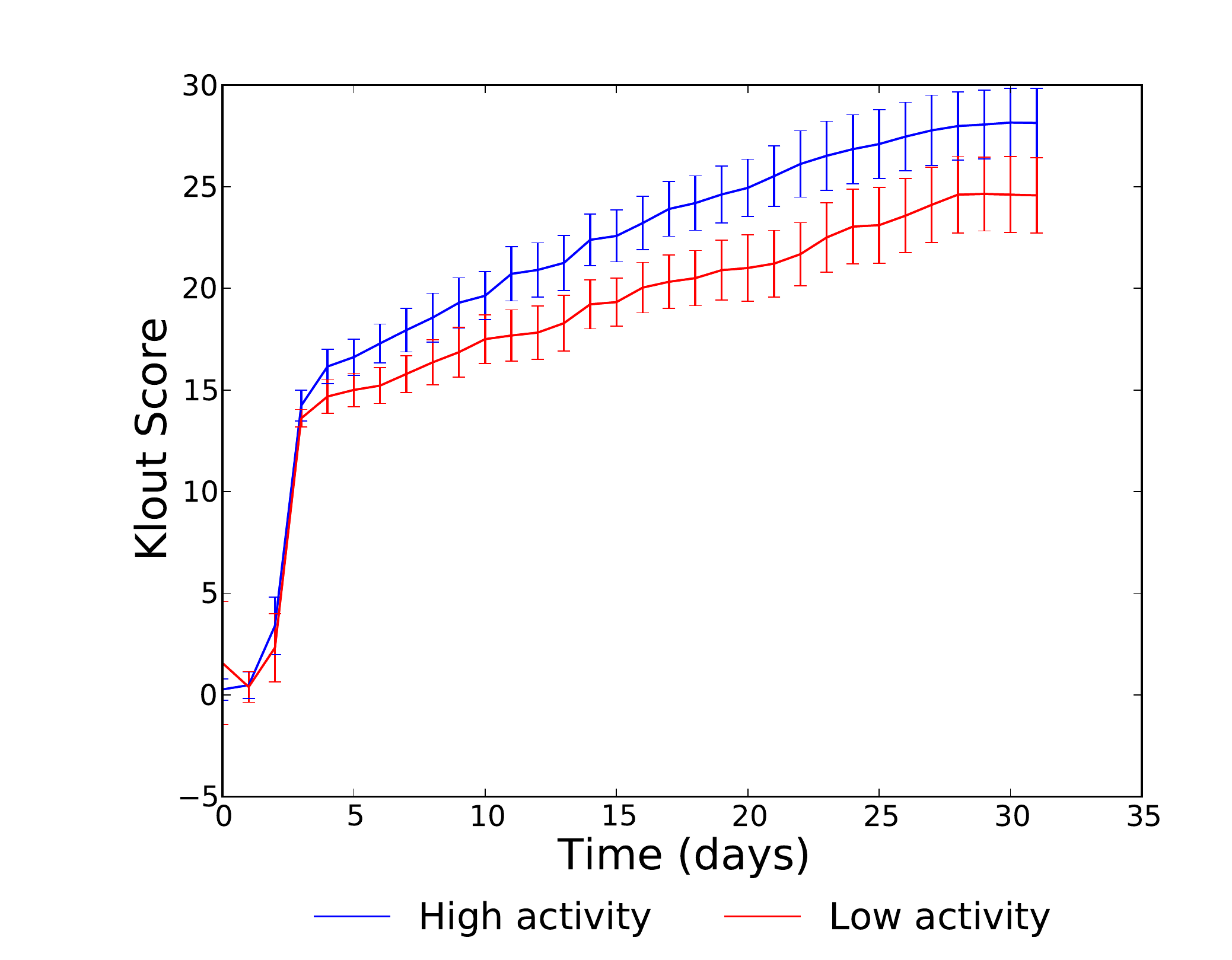}}
  \subfloat[{\bf Message interactions}]{\includegraphics[width=.33\textwidth, height=3.5cm]{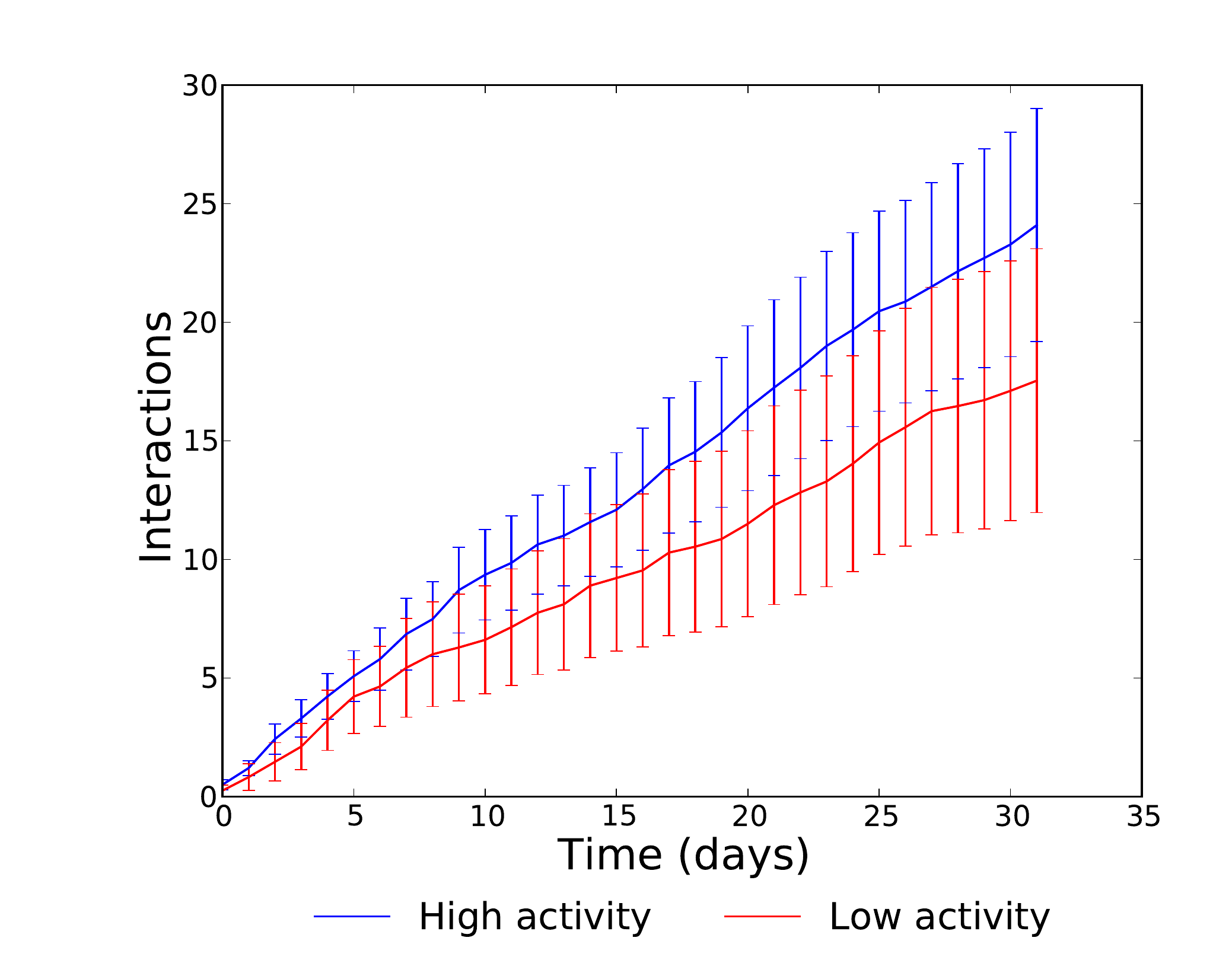}}
\vspace*{-2mm}
  \caption{Infiltration performance of socialbots having different activity levels: (i)~mean number of followers, (ii)~mean Klout score, and (iii)~mean number of message-based interactions with other users.}
  \label{fig:performance-activity}
\vspace*{-3mm}
\end{figure*}


We next study the impact of the socialbots' activity levels,
which we define as lower or higher based on 
how frequently a socialbot posts tweets and follows users.

Figure~\ref{fig:performance-activity}(a) and (b) 
respectively show the mean
number of followers and mean Klout scores of the socialbots
(having the two different levels of activity)
on each day during the experiment. We can see that
socialbots with higher activity levels achieve significantly more
popularity and Klout score than less active socialbots.  
Figure~\ref{fig:performance-activity}(c)
show the mean number of message-based interactions of socialbots 
with other users in Twitter. 
Again, the more active socialbots achieved much more interactions. 

Thus, we find that the more active are the bots, the more
likely they are to be successful in infiltration tasks, as well as in
gaining popularity in the social network.
This is expected, since the more active a bot is,
the higher is the likelihood of its being visible to other users.  
However, it must also be noted that the more active a bot is,
the more likely it is to be detected by Twitter's 
defense mechanisms. 



\subsection{Tweet generating method}

\begin{figure*}[tb]
  \centering
  \hspace*{-5mm}
  \subfloat[{\bf Number of followers}]{\includegraphics[width=.33\textwidth, height=3.5cm]{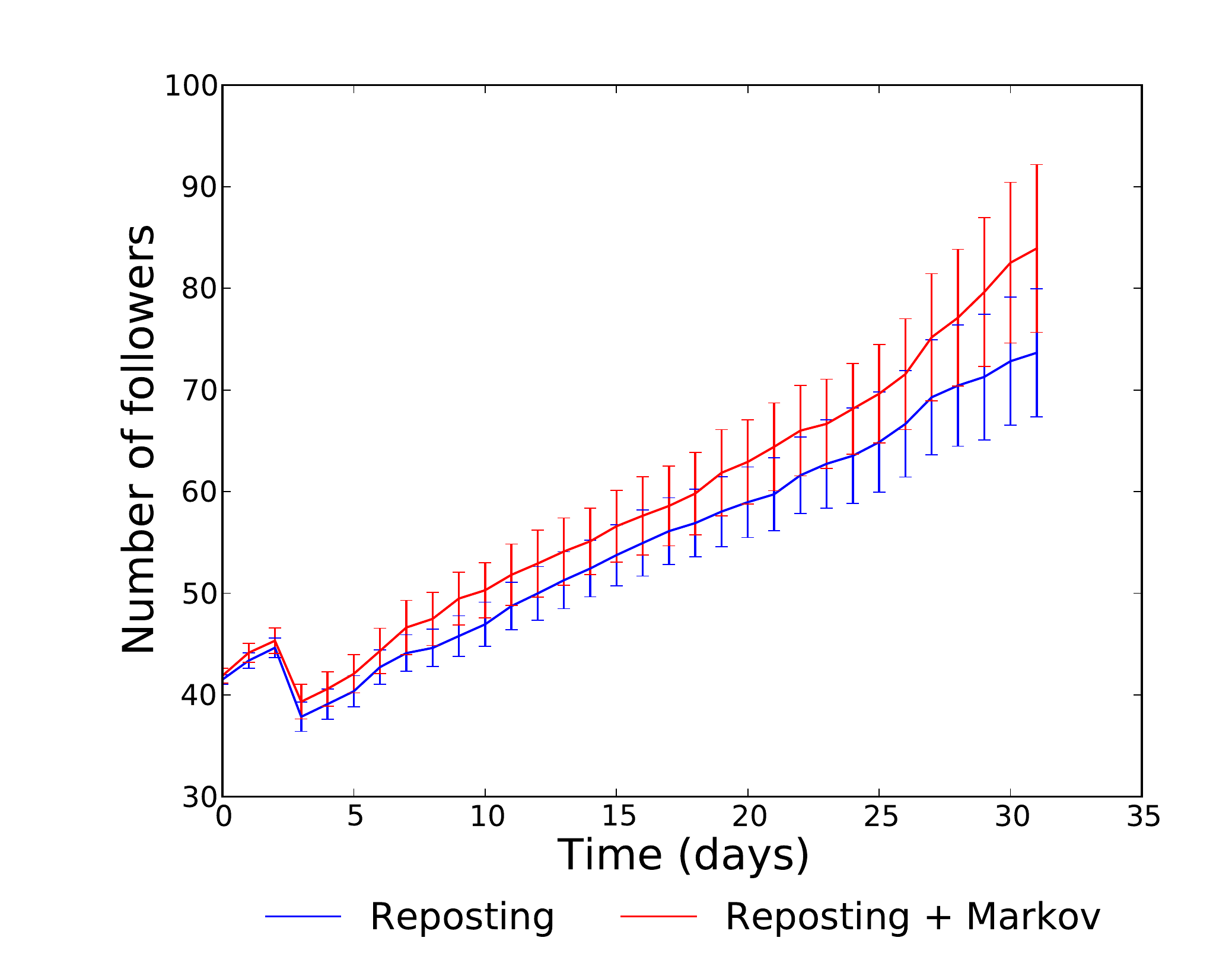}}
  \subfloat[{\bf Klout Score}]{\includegraphics[width=.33\textwidth, height=3.5cm]{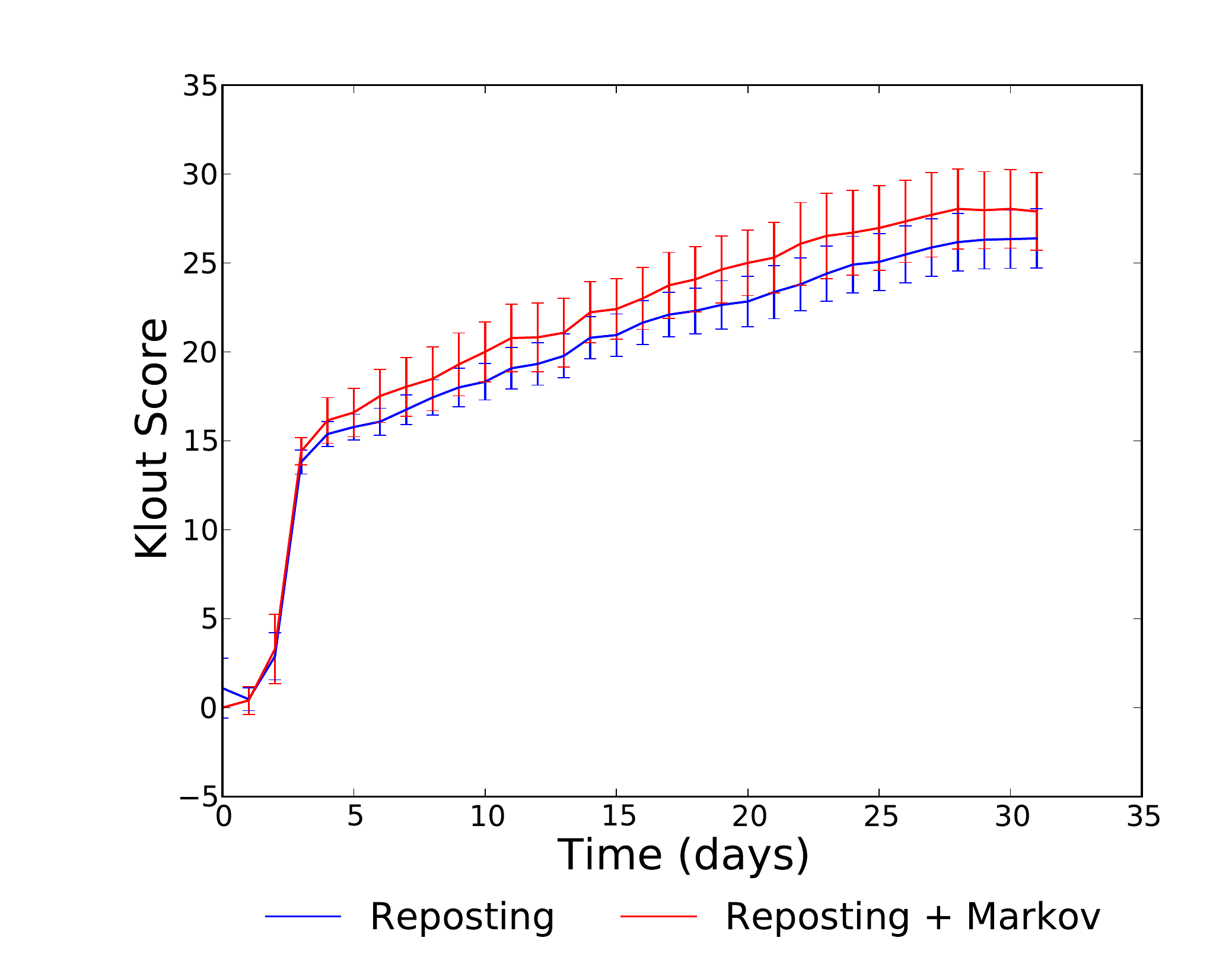}}
  \subfloat[{\bf Message interactions}]{\includegraphics[width=.33\textwidth, height=3.5cm]{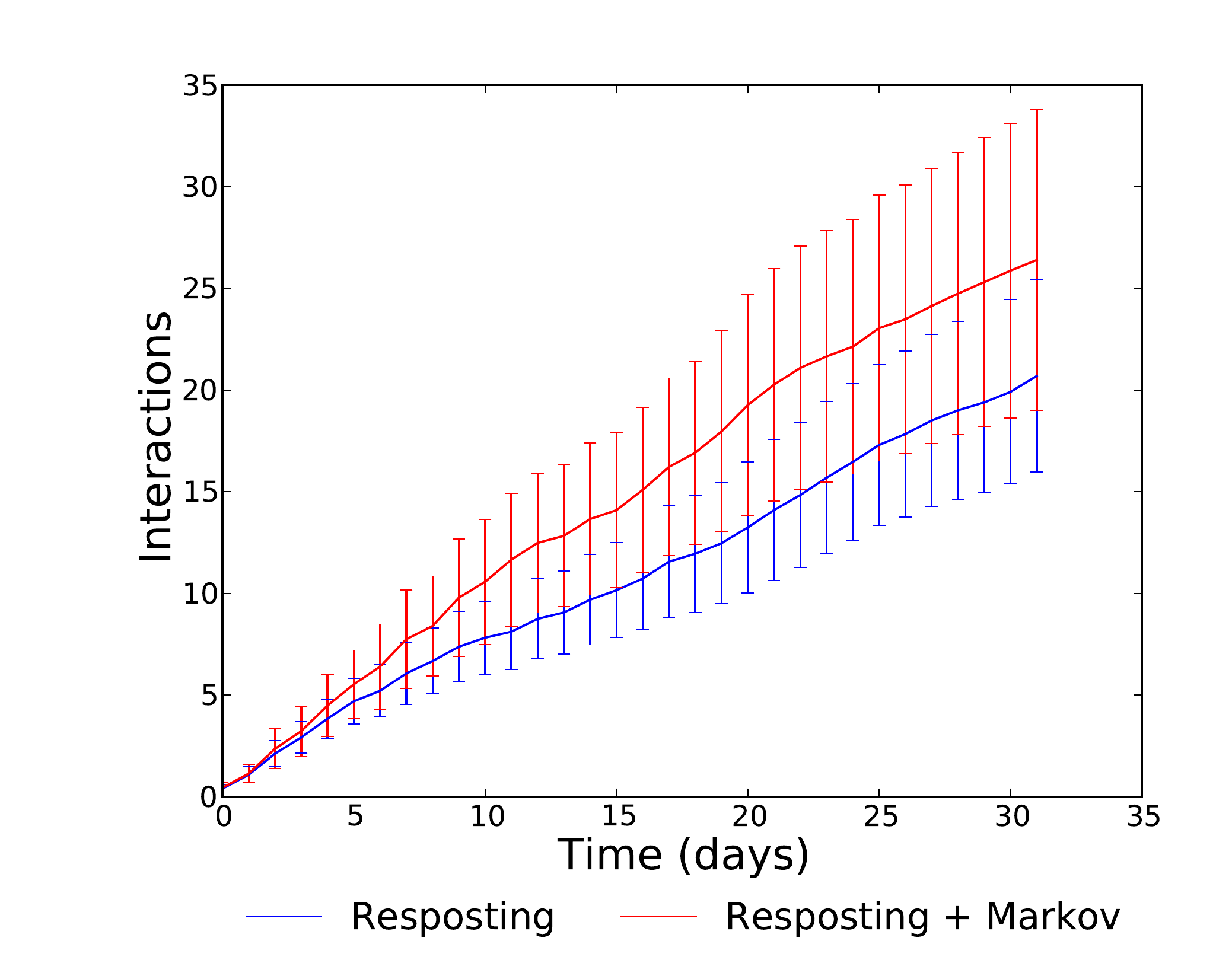}}
\vspace*{-2mm}
  \caption{Infiltration performance of socialbots employing different methodologies to generate tweets: (i)~mean number of followers acquired, (ii)~mean Klout score, and (iii)~mean number of message-based interactions with other users.}
  \label{fig:performance-tweetgen}
\vspace*{-5mm}
\end{figure*}

We next analyze the impact of the tweet generating method used by the socialbots.
Recall from Section~\ref{sec:methodology} that half of our socialbots
only re-posted tweets written by other users (strategy
denoted as `reposting'),
while the other half re-posted tweets 
as well as synthetically generated tweets using a Markov
generator, with equal probability (strategy denoted as `reposting + Markov').

Figure~\ref{fig:performance-tweetgen}(a), (b), and (c) 
respectively show the mean number of followers, mean Klout
scores, and the mean number of message-based interactions acquired by
the socialbots employing the two posting strategies
(on each day during the experiment). 
It is seen that the 
socialbots employing the `reposting + Markov' strategy
acquired marginally higher levels of popularity (number of followers and 
Klout scores), and much higher amount of interactions (social engagement)
with other users. 



The fact that socialbots which automatically 
generated about half of their tweets
achieved higher social engagement
is surprising, since it indicates that users in Twitter
are not able to distinguish between (accounts which post)
human-generated tweets and 
automatically generated tweets using simple statistical models.
This is possibly because a large fraction of tweets in Twitter
are written in an informal, grammatically 
incoherent style~\cite{Kouloumpis:2011}, so that
even simple statistical models can produce tweets with quality similar
to those posted by humans in Twitter. 



\subsection{Target users}

\begin{figure*}
  \centering
  \hspace*{-5mm}
  \subfloat[{\bf Number of followers}]{\includegraphics[width=.33\textwidth, height=3.5cm]{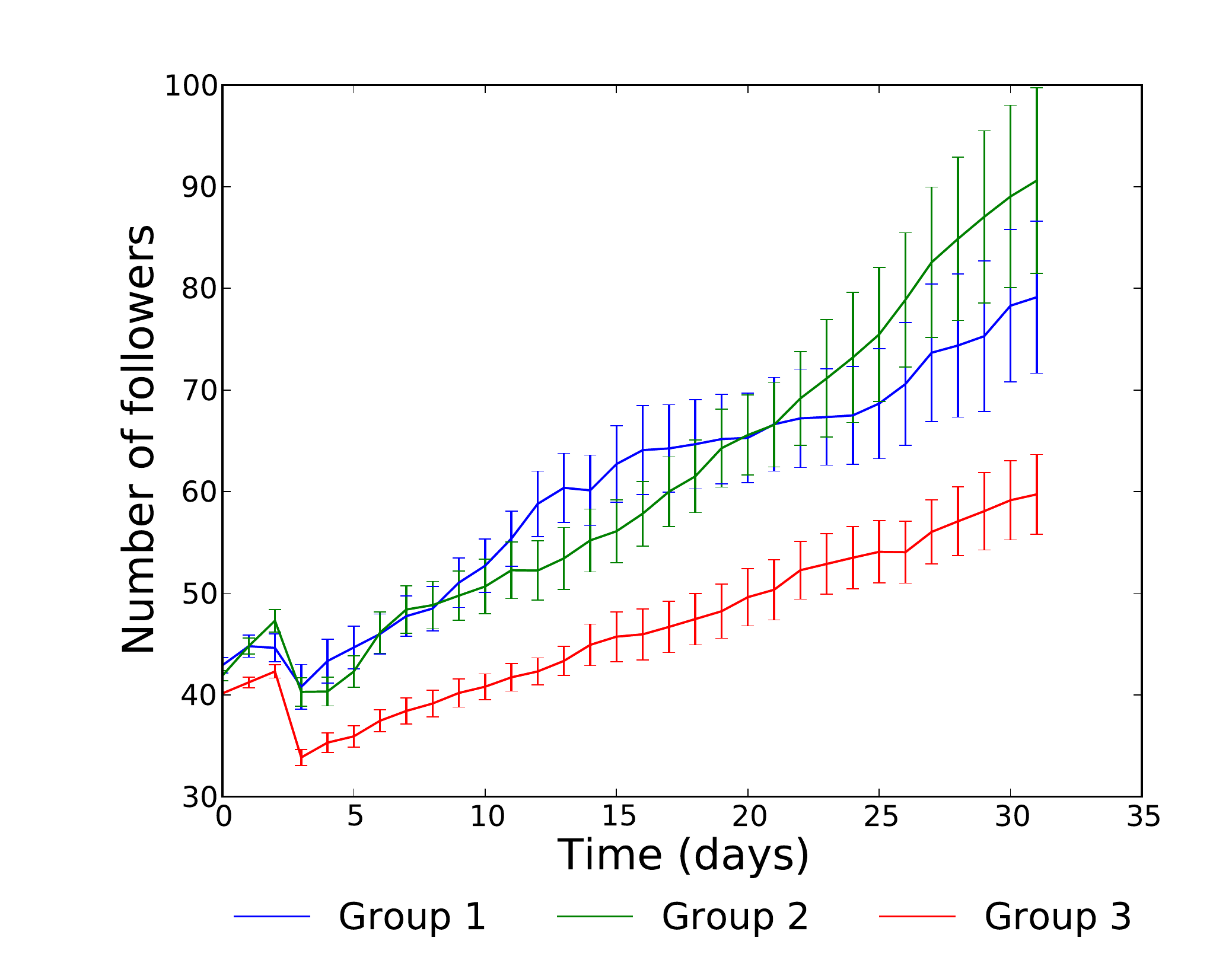}}
  \subfloat[{\bf Klout Score}]{\includegraphics[width=.33\textwidth, height=3.5cm]{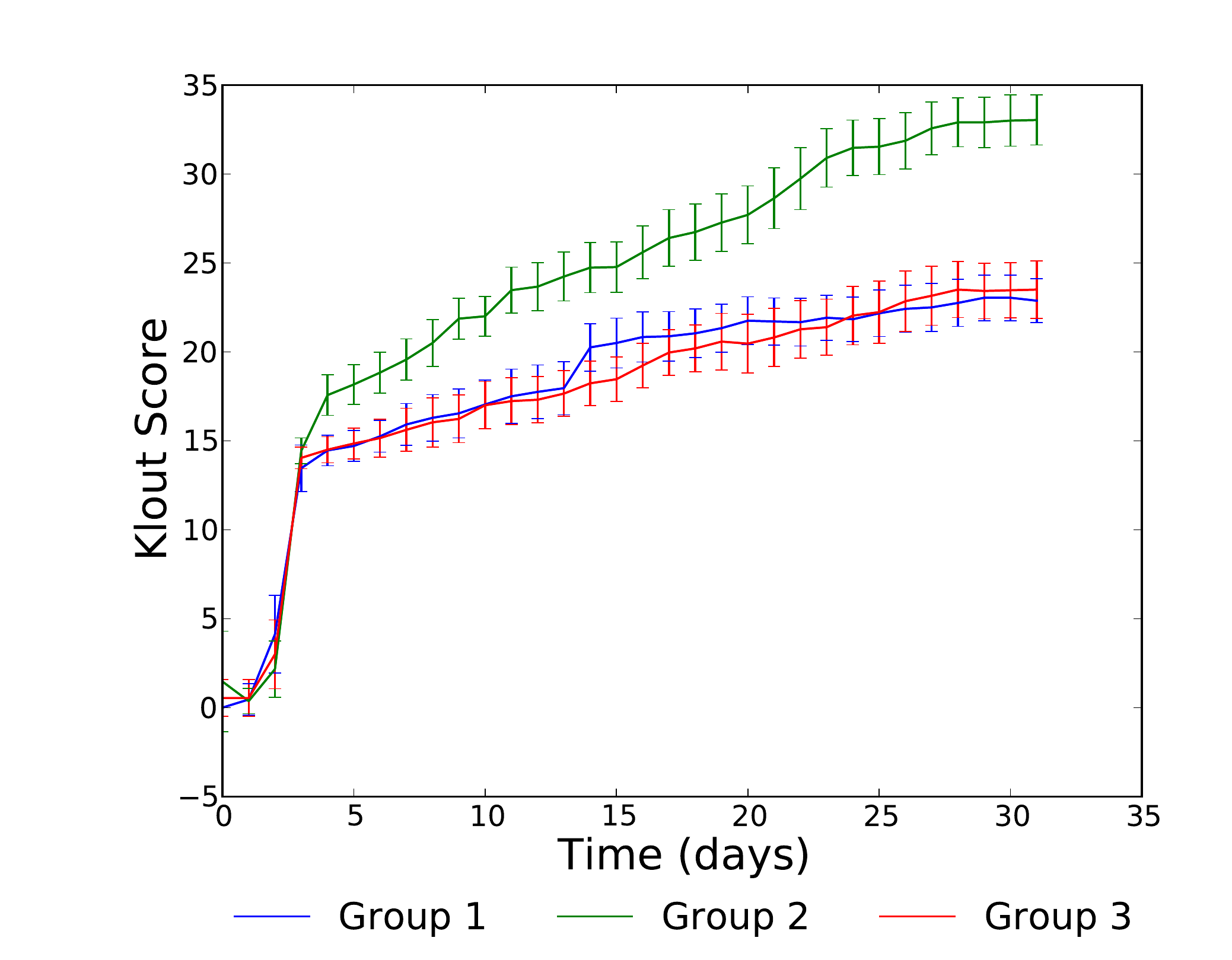}}
  \subfloat[{\bf Message interactions}]{\includegraphics[width=.33\textwidth, height=3.5cm]{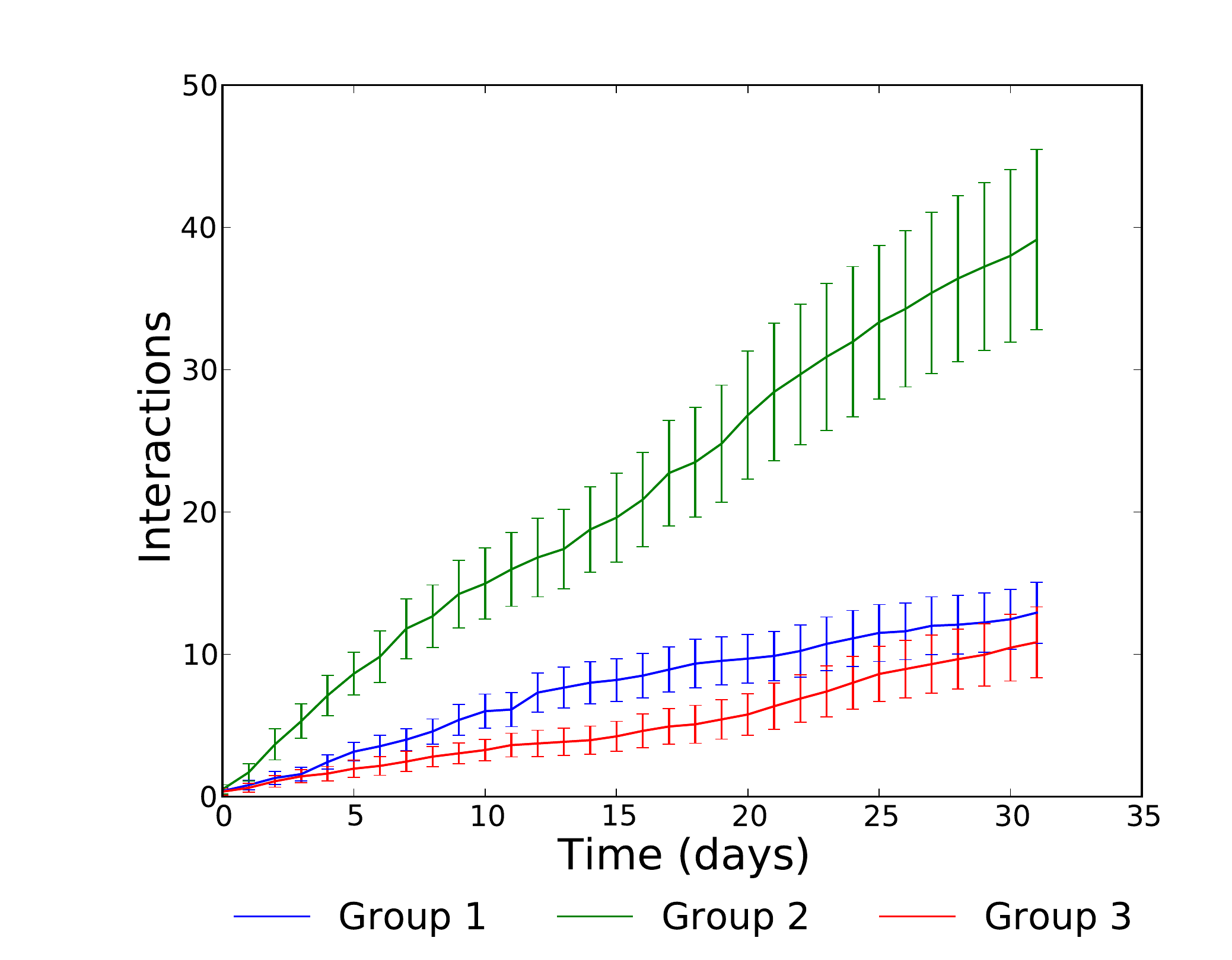}}
\vspace*{-2mm}
  \caption{Infiltration performance of socialbots which followed different sets of target users: (i)~mean umber of followers acquired, (ii)~mean Klout score, and (iii)~mean number of message-based interactions with other users.}
  \label{fig:performance-targetusers}
\vspace*{-5mm}
\end{figure*}

Finally, we analyze the infiltration performance of socialbots
who were assigned different sets of target-users to follow.
Recall from Section~\ref{sec:methodology} that the socialbots
were divided into three groups based on the target-set --
Group 1 followed users selected at random,
Group 2 followed target-users who post tweets on a specific topic
(related to software development),
and Group 3 of socialbots followed target-users who not only post tweets
on the specific topic but are also socially well-connected among themselves.

Figure~\ref{fig:performance-targetusers}(a)
shows the average number of followers acquired by  
each group of socialbots throughout the experiment. 
It is seen that the socialbots in Group~3 had the
lowest number of followers, while those in Group~2 had a significantly 
higher number of followers.
Figure~\ref{fig:performance-targetusers}(b) 
shows the average values of Klout score achieved by our socialbots over time. 
Again, the socialbots in Group~2 have
the highest Klout scores, while the other groups have a similar
performance.
Figure~\ref{fig:performance-targetusers}(c)
shows the average number of
message-based interactions of each group of socialbots (with other Twitter users) over time.
Again, we find that socialbots in Group~2 got significantly more interactions
with other users, and those in Group~3 got the least interactions.

These observations lead to some interesting insights.
Following a set of users who post tweets on a specific
common topic (e.g., software development)
is a more promising approach than following random users (as done by Group 1).
However, although the target-users for both 
Group~2 and Group~3 post tweets on a common topic,
the socialbots in Group~2 achieved significantly higher popularity
and engagement -- this implies that 
infiltrating into {\it inter-connected} groups
of users (Group~3) is far more difficult than engaging with users
without any relation among themselves (Group~2).
Note that this observations differs from those made by similar 
research on Facebook~\cite{Elyashar:2013}.
where it was found that socialbots can effectively infiltrate 
social networks among members of specific organizations.

~\\
\noindent\textbf{Summary:}
The analysis of the impact of various attributes
on the infiltration performance of socialbots shows that,
while certain attributes -- such as
the gender mentioned in the profile -- 
do {\it not} significantly affect
infiltration performance, other attributes
such as the activity level and the choice of the target users
has large impact upon infiltration performance.\\
In this section, we focused on each attribute individually, and
compared the infiltration performance for different 
choices (strategies) of that attribute.
However, we did not attempt to analyze the {\it relative importance
of different attributes}; for instance, we did not investigate which of 
the two attributes (i)~activity level and (ii)~choice of target users
has the larger impact upon infiltration performance.
The next section quantifies the relative impact of the various attributes
on infiltration performances.

\section{Assessing Attribute Importance}\label{sec:attributes}

In this section, we aim to assess the relative importance of the 
different attributes and infiltration strategies of the socialbots.
Our objective is to {\it quantify} which strategy (or, combination of strategies)
has the greatest impact in deciding how the socialbots can infiltrate
specific groups of target users. 
Note that, differently from Section~\ref{sec:performance}, here
we consider the socialbots' performance in infiltrating specific
groups of targeted users.

We present a {\it factorial design experiment} to assess the relative impact of the different infiltration strategies. We begin by briefly describing
how we designed our experiments, and then discuss the obtained results.  

\subsection{$2^{k}$ factorial experiment}

We here include a brief description of the theory of $2^{k}$ factorial experiments;
we refer the reader to~\cite{jain} for a comprehensive description.

An experimental design strategy is usually necessary in scenarios with a large number of factors, as an attempt to reduce the number of factors that will be part of the experiment.
Particularly, $2^k$ designs refer to experimental designs with $k$ factors where each factor has the minimal number of levels, just two. As an illustrative example, suppose an
experimental performance scenario in which three factors -- memory, disk, and CPU of a machine -- can potentially affect the performance of an algorithm. 
Suppose now that each
experiment takes about one day to run and there are 10 possible types of memory, 10 types of disks, and 10 types of CPUs to be tested. Running an experiment with all
possibilities would take $10 \times 10 \times 10 \, = \, 1,000$ days.
Instead of running all possibilities, a $2^k$ design would consider two (usually extreme) types of memory, two types of disk, and two types
of CPUs to compare, which would result in only $2^3 \, = \, 8$ days of experiments. 
The theory of factorial experiments~\cite{jain}
 would then allow one to estimate how much each 
factor impacts on the final result, a key information
to help decide on which factors an experiment should focus. 

Note that, differently of the above example, our goal here is {\it not} primarily to reduce the number of experiment scenarios.
Instead we use a $2^k$ design to infer how much a factor -- 
which, in our case, correspond to attributes 
like gender, activity level, and posting method --  
impacts the different infiltration metrics.

\subsection{Factorial experiment on the socialbot infiltration} 

For certain applications, 
the objective of socialbots might be to infiltrate a particular
target group of users. Hence, we here individually consider the success
of our socialbots in infiltrating each of the three target groups
(which were described in Section~\ref{sec:methodology}).
For each target group, we consider the three infiltration
metrics stated earlier -- the number of followers acquired, the 
number of message-based interactions 
and the Klout score.
Then, for each metric and each target group, we executed a $2^3$ design
considering the attributes and their values as described in Table~\ref{table:factors}, resulting in $3 \times 3 \times 2^3 = 216$ experiments.  We performed experiments that associates $+1$ or
$-1$ for the strategies employed for each attribute. All experimental configurations for all datasets were averaged over 5 results, which is the number of socialbots
in each configuration. 
 
\begin{table}[tb]
\center
\small
  \begin{tabular}{lll} \hline
  \multicolumn{1}{c}{Factor} & \multicolumn{1}{c}{$-$1} &\multicolumn{1}{c}{$+$1}  \\ \hline
  Gender (\textbf{G})          & Female       & Male          \\
  Activity Level (\textbf{A})  & Low activity & High activity \\
  Posting Method (\textbf{P})  & Repost       & Repost+Markov \\ \hline
  \end{tabular}
  \caption{Factors used in the factorial experiment for the socialbot infiltration study.}
  \label{table:factors}
\end{table}

The basic idea of the factorial design model consists of
formulating $y$, the infiltration impact, 
as a function of a number of factors and their possible combinations,
as defined by Equation~\ref{eqn:factorial}.  
Here, GP, AP, AG, and GAP account for all possible combinations among the factors. 
For instance, the experiments for `GP' attempts to measure
the impact of a certain combination of the attributes Gender (G) and Posting method (P) (e.g., `Female and Repost', or `Male and Repost+Markov').
\begin{equation}
y \; = \; Q_0 \, + \, \sum_{i \in F} Q_i \cdot x_i
\label{eqn:factorial}
\end{equation}
where $F = \{G, A, P, GA, GP, AP, GAP\}$ and $x_i$ is defined as follows.\\
\noindent
$
x_{G} = \left\{\begin{tabular}{cl}
-1 & \mbox{if Female} \\ 
+1 & \mbox{if Male}
\end{tabular}\right.
$\\
$
x_{A} = \left\{\begin{tabular}{cl}
-1 & \mbox{if Low activity} \\ 
+1 & \mbox{if High activity}
\end{tabular}\right.
$\\
$
x_{P} = \left\{\begin{tabular}{cl}
-1 & \mbox{if Repost} \\ 
+1 & \mbox{if Repost + Markov}
\end{tabular}\right.
$\\

\noindent and the $x_i$'s for the feature-combinations (e.g., AG, GP) are
defined from the values of $x_G$, $x_A$, and $x_P$ following 
the standard way described in~\cite{jain} (details omitted for brevity).

In the above equation, $Q_i$ is the infiltration performance
(according to a certain metric like number of followers, or Klout score)
when strategy $i \in F$ is applied, and 
$Q_0$ stands for the average infiltration performance, 
averaged over all possible features and their combinations.
By empirically measuring $y$ according to different 
feature-combinations (which, in our case, refer to the various socialbot strategies), 
we can estimate the values of the different $Q_i$ and $Q_0$.
This allows us to understand by how much each 
factor impacts the final infiltration performance.

Instead of presenting results for all possible values of $Q_i$, 
we focus on the {\it variations of $Q_i$ due to changes in the features
(or their combinations)}, which helps to estimate
the importance of a particular factor to the final result.  
As an example, if we find that a factor accounts 
for only 1\% of total variation on the results, we can 
infer that this attribute is unimportant for
infiltrating Twitter with a socialbot.  

As proposed in~\cite{jain}, the importance of the various factors
 can be quantitatively estimated by assessing the 
{\it proportion of the total variation in the final result 
that is explained by each factor}. 
To compute this variation, we first consider the variation of $y$ 
(as defined by Equation~\ref{eqn:factorial})
across all runs, and then compute $SS_T$ as the sum of 
the squared difference between each measured value of $y$ 
and the mean value of $y$.  Then, we compute $SS_i$ as 
the {\it variation only due to the changes on factor $i$}, which
can be computed similarly to $SS_T$, but considering only those runs
in which the values of the factor $i$ were changed.  
Finally, we calculate the fraction of variation due to factor $i$ as
$\frac{SS_i}{SS_T}$.  
We now use this metric to compute the impact of each attribute 
for different infiltration metrics and groups of target users.

\begin{table*}[tb]
\center
\small
\begin{minipage}{0.8\textwidth}{
  \begin{tabular}{l|ccccccc}
            & Gender (G) & Activity level (A) & Posting method (P) & GA   & GP    & AP             & GAP  \\ \hline	      
    Group 1 & 7.43       & \textbf{53.75}     & 12.44              & 5.20 & 0.85  & \textbf{20.10} & 0.23 \\
    Group 2 & 3.99       & \textbf{72.65}     & 2.77               & 4.38 & 3.53  & 2.81           & 9.87 \\    
    Group 3 & \textbf{20.52}      & \textbf{49.27}     & 2.02      & 2.40 & 5.42  & 12.71          & 7.66 \\ \hline
  \end{tabular}
  \caption{Percentage variation in the number of followers explained by each kind of attribute}
  \vspace{3mm}
  \label{table:followvariance}
}
\end{minipage}
\medskip
\begin{minipage}{0.8\textwidth}{
  \begin{tabular}{l|ccccccc}
	  & Gender (G)     & Activity level (A) & Posting method (P) & GA    & GP             & AP             & GAP   \\ \hline
  Group 1 & 0.03           & \textbf{36.58}     & 13.87              & 0.31  & 2.83           & \textbf{44.74} & 1.64 \\
  Group 2 & 0.00           & \textbf{40.56}     & 7.26               & 20.67 & 19.39          & 6.34           & 5.77 \\  
  Group 3 & 12.71 	    & \textbf{43.23}     & 4.51               & 19.60 & 8.18           & 1.19           & 10.58 \\ \hline
  \end{tabular}
  \caption{Percentage variation in the number of message-based interactions explained by each kind of attribute}
  \label{table:engagementvariance}
}
\end{minipage}
\medskip
\begin{minipage}[c]{0.8\textwidth}{
  \begin{tabular}{l|ccccccc}
	  & Gender (G) & Activity level (A) & Posting method (P) & GA    & GP    & AP    & GAP   \\ \hline
  Group 1 & 0.46       & \textbf{41.32}     & 21.69              & 0.00  & 0.61  & \textbf{35.90} & 0.02 \\
  Group 2 & 7.58       & \textbf{31.98}     & 12.62              & 15.93 & 15.93 & 10.19 & 5.78  \\
  Group 3 & 12.58      & \textbf{31.42}     & 17.92              & 12.94 & 12.37 & 2.13  & 10.65 \\ \hline
  \end{tabular}
  \caption{Percentage variation in the Klout score explained by each kind of attribute}
  \label{table:kloutvariance}
}
\end{minipage}
\end{table*}

\subsection{Attribute Importance}

We begin by analyzing to what extent each of the attributes 
impacts {\it the number of followers} acquired by the socialbots.  
Table~\ref{table:followvariance} shows the
variation explained by each attribute in the number of followers acquired by the socialbots from each of the target groups.  
We note that the {\it activity level} of a socialbot 
is the most important attribute for Group~1
(random users) of target users, being 53.75\% responsible 
for deciding the number of followers acquired by a socialbot. 
The second most important attribute is the {\it posting method}
(i.e., technique used to generate the tweets),
which accounts for 12.44\%  of the variation on the number of followers.  
The combination of these two attributes 
(AP column in Table~\ref{table:followvariance}) leads to a high
variation (about 20\%) on the number of followers as well.  

Similar observations can be made from Table~\ref{table:engagementvariance} 
and Table~\ref{table:kloutvariance}, which shows 
the percentage variation explained by each attribute on the
number of message-based interactions (i.e., number of tweets retweeted or favorited, 
number of mentions, and number of replies) and on the Klout Scores, respectively.

We also observe that the importance of some of the attributes 
varies significantly according to the group of users targeted
by the socialbots. For instance, the {\it gender} attribute
has a great importance in the experiments with target users 
from Group~3, being responsible for 
20.52\% of the variation of the number of followers 
(Table~\ref{table:followvariance})
and 12.71\% of the message-based 
interactions (Table~\ref{table:engagementvariance}) 
when the target users are from this group.\footnote{We found
that the users in Group~3 were more likely to follow and interact with
socialbots having female profiles.}
However, the gender attribute does not seem to have much influence on the other target groups.
This suggests that the gender of the socialbots can make a 
difference if the target users are gender-biased, or susceptible to follow and
interact with users of a certain gender.

%

\section{Concluding discussion}

In this work, we exposed Twitter's vulnerability against large-scale socialbot attacks that can affect both Twitter itself and services built on crowd-sourced data gathered from
Twitter.  This problem is a clear adversarial fight, or as usually called, a cat and mouse fight.  We put ourselves in the mouse's shoes (i.e., assumed the perspective of
socialbot-developers) as an attempt to bring to the research community a novel perspective to the problem.  We analyzed the importance of various attributes in deciding the
infiltration performance of socialbots, and designed a $2^k$ factorial design experiment in order to  quantify the extent to which different socialbot strategies impact their
infiltration performance.

We believe that our findings have a number of implications for designers of
spam defense mechanisms in OSNs.  
First, we demonstrated that it is possible to infiltrate 
Twitter with simple automated strategies -- 
this calls for better defense mechanisms than those that
are currently deployed on Twitter. 
We also show that existing influence metrics such as Klout are 
vulnerable to socialbot strategies.
Second,  malicious activities have different nuances in the Twitter platform, 
including Sybil attacks~\cite{Viswanath:2010}, link-farming~\cite{ghosh:2012}, 
content pollution~\cite{Lee:2011}, content credibility~\cite{Castillo:2011:ICT:1963405.1963500}, search SPAM~\cite{benevenuto@ceas10}, phishing~\cite{Benevenuto:2011}, and so on.
Socialbots may be used to perform any of these forms of attacks;
moreover, their strategies might change depending on the particular set
of users they want to infiltrate or influence. 
Hence, our efforts towards understanding the best socialbot strategies 
may suggest improvements for different defense mechanisms 
designed to counter specific types of malicious activities in Twitter. 
Third, our study suggests some limitations of existing mechanisms for detection
of bots / fake accounts, and potential ways to improve them:\\
(i)~We find that re-posting others' tweets is a simple but promising
strategy for socialbots, especially considering that existing Twitter
defenses could not detect most of the bots employing this strategy. This suggests
that those accounts for which a large fraction of tweets are re-posts / retweets can be monitored.\\
(ii)~We see that Twitter users, in general, are {\it not} good at distinguishing
tweets posted by real users and tweets generated automatically using statistical models.
This implies that relying on user-generated reports for identifying
bots / fake accounts (as done by Twitter~\cite{twitter-shut-spammers})
may not be a promising approach. Rather, approaches that investigate the linguistic 
aspects of tweets could be explored to identify automatically generated tweets.\\
(iii)~It is, in a way, comforting to note that, in order to achieve high infiltration success
in a short time, socialbot accounts need to be highly active, 
e.g., they need to post tweets and follow users almost every hour. 
This implies that it might be sufficient to monitor those accounts which
are regularly active, in order to prevent bots from becoming influential.



As future work, we note that most socialbot attacks ultimately aim to manipulate public opinion about political candidates, themes or products.  Although there exists evidence of
opinion change~\cite{Muchnik:2013,Timberg:2013:Online}, it is still unclear  whether, or to what extent, socialbots can influence people's decisions about products, brands, or
political candidates.  This is an open research topic that we plan to investigate in future.

\section*{Acknowledgements}

This work was supported by grants of CAPES, CNPq, and Fapemig.

\bibliographystyle{abbrv}

\end{document}